\documentclass[aps,superscriptaddress,nofootinbib,eqsecnum,pra,notitlepage,twocolumn]{revtex4-2} 

\usepackage{amssymb}
\usepackage{amsmath,amsfonts,amsthm,bm}
\usepackage{graphicx,xcolor}
\usepackage{float}
\usepackage{hyperref}
\usepackage{subfigure}
\usepackage{dcolumn}
\usepackage{soul}
\usepackage{ulem}
\usepackage{verbatim}

\DeclareMathAlphabet{\mathcal}{OT1}{pzc}{m}{it}

\begin{document}

\title{Confinement-induced unatomic trimer states in mass-imbalanced systems}

  \author{R. M. Francisco}

\affiliation{Instituto Tecnol\'{o}gico de Aeron\'{a}utica, DCTA,
  12228-900 S\~{a}o Jos\'{e} dos Campos, SP, Brazil}

\author{D. S. Rosa}

\affiliation{Instituto Tecnol\'{o}gico de Aeron\'{a}utica, DCTA,
  12228-900 S\~{a}o Jos\'{e} dos Campos, SP, Brazil}

\affiliation{Université Paris-Saclay, CNRS/IN2P3, IJCLab, 91405 Orsay, France}
  
  \author{T. Frederico}

\affiliation{Instituto Tecnol\'{o}gico de Aeron\'{a}utica, DCTA,
  12228-900 S\~{a}o Jos\'{e} dos Campos, SP, Brazil}

\author{G. Krein}

\affiliation{Instituto de F\'isica Te\'orica, Universidade Estadual Paulista,
Rua Dr. Bento Teobaldo Ferraz, 271-Bloco II, 01140-070 S\~ao Paulo, SP, Brazil}

\author{M. T. Yamashita}

\affiliation{Instituto de F\'isica Te\'orica, Universidade Estadual Paulista,
Rua Dr. Bento Teobaldo Ferraz, 271-Bloco II, 01140-070 S\~ao Paulo, SP, Brazil}

\begin{abstract}
As resonantly interacting trimers of the type AAB are progressively squeezed from $D=3$ to $D=2$, unatomic states emerge. We calculated the contacts from the high momentum tail of the single particle densities. The sharp increase of the contacts serves as a signature of the transition between the Efimov and unatomic regimes, characterized by the emergence of continuous scale invariance when the system reaches a critical dimension, $D_c$. This continuous scale invariance starts to dominate the behavior of the system at the dimension $\overline{D}<D_c$, below which the trimers momentum distribution tails exhibit a power-law behavior signaling the unatomic regime. To illustrate our findings, we studied compounds of the forms $^{7}$Li$-^{23}$Na$_{2}$, $^{7}$Li$-^{87}$Rb$_{2}$ and $^{7}$Li$-^{133}$Cs$_{2}$. The increase in the mass-imbalance of the trimers reduces the interval between $D_c$ and $\overline{D}$. The emergence of unatomic states can be experimentally verified by observing the two-body contact parameter, which is a quantity directly related to the thermodynamic properties of the gas.
\end{abstract}
\maketitle

\section{Introduction}

In 2007, Georgi~\cite{georgi} hypothesized that, beyond the Standard
Model (SM) of elementary particles, there exists a sector of hidden symmetries, referred to as unparticles, which are generated by local operators with correlation functions that exhibit power-law behavior. The unparticles can only be observed indirectly by analyzing the recoil momentum distribution of SM particles that are produced in association with them. Analogously to the relativistic unparticle, Hammer and Son examined a field of a nonrelativistic conformal field theory (CFT)~\cite{mehenCFT,nishidaCFT}, which they  named unnucleus~
\cite{commentHammerSon,HammerSon}. They argued that the unnucleus, which is characterized by a mass and a nonrelativistic conformal dimension, exists in nature and can be realized in nuclear reactions involving neutron emission~\cite{unnucleus1,unnucleus2,unnucleus3}, occurring when momentum~$p$ is transferred with a length ${\hbar}/p$ lying between the range of the nuclear forces
and the two-neutron scattering length.

Extending the definition of the unparticle and unnucleus to the context of cold atoms, it is possible to realize unatomic systems by driving the few-body interactions close to the unitarity limit. In this limit, where the scattering length is infinite, three-boson bound states are generated by a scale-invariant Hamiltonian, which gives rise to the well-known Efimov effect~\cite{efimov0,efimov1} (for reviews,
see Refs.~\cite{Braaten:2004rn,Naidon:2016dpf,Greene:2017cik,Hammer:2019poc}). In principle, the Efimov states, which exhibits discrete scale symmetry with a log-periodic behavior, would not allow the existence of unatomic states, which are characterized by a continuous scale symmetry with a power law behavior. Because of this, until recently, the observation of continuous scale symmetries was only proposed for fermionic systems~\cite{fermions1,unnucleousatomic}. By using the method of noninteger dimensions to mimic the deformation of the trap that confines cold atoms~\cite{D1,D2,D3,D4,D5}, in Ref.~\cite{unatomic}, it was demonstrated that unatomic states can be realized in bosonic systems in the unitarity limit by continuously squeezing it from $D=3$ to $D=2$, reaching a critical dimension in which the discrete scaling symmetry disappears and a continuous one takes place. 

To establish a connection between deformations in the geometry of the traps that confine atomic systems and the formalism of noninteger dimensions, the aspect ratio of the confining trap can be related to an effective dimension $D$ by means of the equation ~\cite{garridoconection3,garridoconection4}:
\begin{equation}
\frac{3(D-2)}{(3-D)(D-1)}= \frac{b_{ho}^2}{r_{2D}^2}\,,
\label{eq:Dtrap}
\end{equation} 
where  $b_{ho}$ is the oscillator length and $r_{2D}$ is the root mean square radius of the three-body system in two dimensions. The findings in Refs.~\cite{garridoconection3,garridoconection4} suggest that this relation is independent of the details of the two-body potential. In practical terms, nowadays it is possible to create effective two-~\cite{BEC2D} and one-dimensional~\cite{BEC1D} setups by compressing atomic clouds, but, as far as we know, it is still a challenge to perform continuous variations in the geometry of the trap while controlling the Feshbach resonance. 

To investigate the transition from the Efimov regime to the unatomic one, in Ref.~\cite{unatomic}, contact parameters were computed in the context of three-boson bound states, where it was demonstrated that by continuously squeezing the trap from $D=3$ to $D=2$, one finds a narrow dimensional interval associated with two consecutive spikes in these parameters: one for the three-body contact, followed by another one for the two-body contact. Contact parameters are important quantities that describe universal relations that were found by Tan while studying the tail of the two-body momentum distribution of unitary Fermi gases~\cite{Tan1,Tan2,Tan3}.
Besides that, the probability of finding three clustered atoms in unitary Bose gases is associated to an additional contact parameter. Measured in a series of contexts (e.g. those of Refs.~\cite{fletcher0,musolino,wild,makotyan}), they parameterize thermodynamic relations between macroscopic observables, such as energy, momentum distribution and response functions of low-temperature gases with short-range interactions. 

In this work, we extend the study of unatomic states in Ref.~\cite{unatomic} to mass-imbalanced three-body systems of the type AAB, being the A particles heavier than the B particle. To illustrate our results, we investigate compounds of the form $^{7}$Li-AA, namely $^{7}$Li$-^{23}$Na$_{2}$, $^{7}$Li$-^{87}$Rb$_{2}$ and $^{7}$Li$-^{133}$Cs$_{2}$, which are systems of current experimental interest~\cite{Juris}. To achieve these goals, we use the formalism of Faddeev wave decomposition alongside the Bethe-Peierls boundary condition. For a detailed discussion about this formalism, we refer to the works in Refs.~\cite{betpeiPRA,Nb_AAB_Ddim_Efimov}. For completeness, the main equations are reviewed in the appendix~\ref{appa}. 

\section{\texorpdfstring{Scale parameter in noninteger dimensions} {$D$-Dimensional} }
\label{section2}

The trimer embedded in a $D$-dimensional geometry is described as an eigenstate of the free Hamiltonian with pairwise contact interaction  under the Bethe-Peierls (BP) boundary condition~\cite{bethe}. Considering the case where all three boson pairs interact resonantly, the BP leads to a characteristic equation~\eqref{BPsystem} determining the scale parameter $s_n$ (for a short review of the method described above, please see App.~\ref{appa}). Ranging between three and two dimensions, one finds that the transcendental equation displays an infinite number of real solutions and only one imaginary, defined as $s_n \equiv \pm s_0$. The infinite number of real solutions are limited by requiring that the spectator function derived from the Faddeev component of the trimer wave function solves the trimer Skorniakov and Ter-Martirosyan equation~\cite{STM} in $D$-dimensions~\cite{dsrSTM}. This procedure was discussed in Ref.~\cite{unatomic} for three identical atoms and is easily extended to mass imbalanced systems. The conclusion is that the real scale parameter, $s_n \equiv s_1$, has to be constrained in the interval $-1<s_1<1$. It is worthwhile to also observe that Eq.~\eqref{radialwavefunc} is symmetric under $s_n \rightarrow -s_n$. From this point, we use $s_1$ in the interval $-1 < s_1 < 0 $ and the positive values $s_0 \geq 0$, where $s_n = is_0$.

\begin{center}
\begin{figure}[h]
\includegraphics[width=8.7cm]{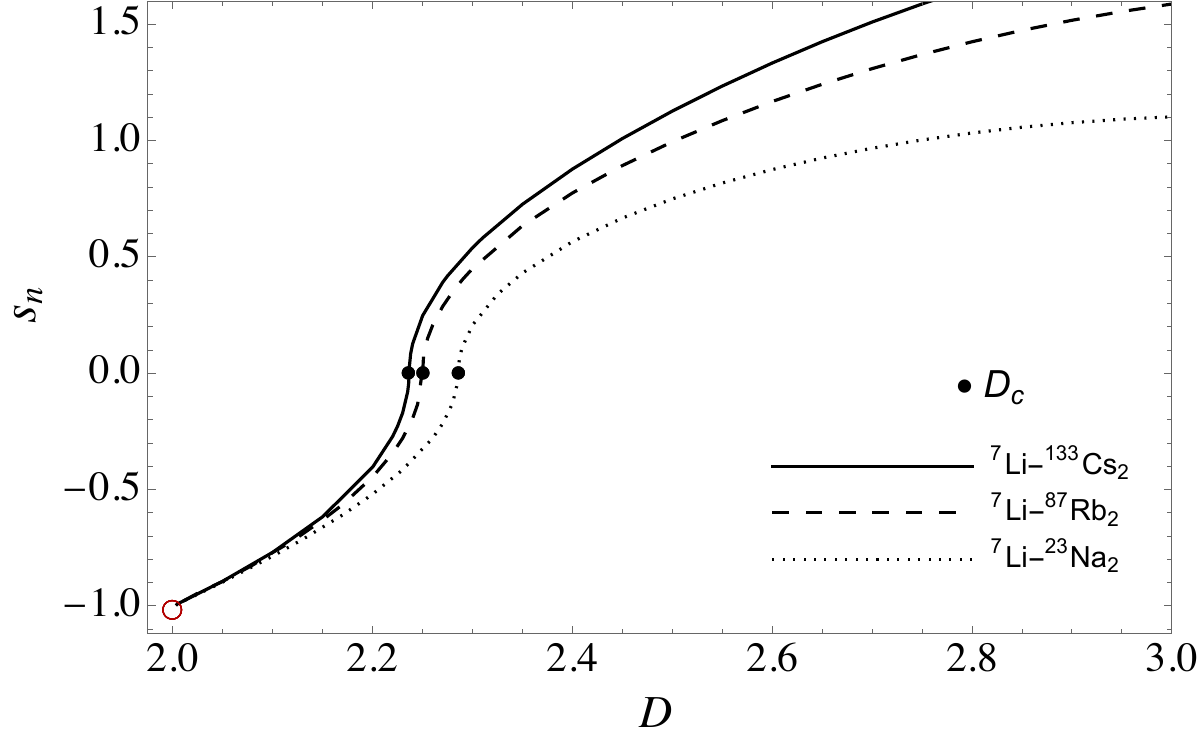}
\caption{Scale parameter solutions $s_n$ ($n=1$ for $s_n<0$ and $n=0$ for $s_n\geq0$) for compounds of the form $^{7}$Li-AA embedded in $D$-dimensions. The black points denote the critical dimensions, which are given by $D_c = 2.2855$, $2.2494$ and $2.2364$ for the $^{7}$Li$-^{23}$Na$_{2}$, $^{7}$Li$-^{87}$Rb$_{2}$, and $^{7}$Li$-^{133}$Cs$_{2}$ compounds, respectively. The empty red cycle corresponds to the limit $D \rightarrow 2_{+}$.}
\label{fig1}
\end{figure}
\end{center}

The $D$-dimensional behavior of the scale parameter is illustrated in Fig.~\ref{fig1} for the compounds $^{7}$Li$-^{23}$Na$_{2}$, $^{7}$Li$-^{87}$Rb$_{2}$ and $^{7}$Li$-^{133}$Cs$_{2}$. The black points denote the critical dimensions ($D_c$) where a transition occurs between the discrete and continuous scale symmetries. Above this point, the physics describing the systems is the Efimov one. For different properties and phenomena in this range of dimensions, please see Refs.\cite{garridocontinum,impuritiesmendes} and references therein. Below the critical dimension, the unparticle nature of the bound three-boson systems is revealed. The empty red circle indicates the $D=2$ limit, where the resonant three-boson systems go to the zero three-body energy limit. In this limit, the energy of the three-body bound state is proportional to the two-body one~\cite{dipole,bellotti2d}. As we can observe in the figure, the greater the mass imbalance between the particles in the compound, the greater the value of the scale parameter. This behavior is well known and it is the one that made it possible to measure excited Efimov states.

%%%%%%%%%%%%%%%%FIGURE01%%%%%%%%%%%%%%%%%%%
%%%%%%%%%%%%%%%%%%%%%%%%%%%%%%%%%%%%%%%%%%%%%

\section{D-dimensional regimes in momentum space}
\label{section2}

In this section, we analyze the behavior of the mass-imbalanced $AAB$ systems in two different regimes, namely Efimov and unatomic. We start by obtaining the shallow state asymptotic behavior of the spectator functions and comparing it to the finite energy case. With the spectator functions, we compute the momentum distributions and show how it behave as the effective dimension is changed. Finally, we calculate the contact parameters, which are universal quantities that can be used to parameterize thermodynamic relations between macroscopic observables, such as the momentum distribution, energy, and response functions of low-temperature gases interacting via short-range interactions~\cite{fletcher0,musolino,wild,makotyan}. 

\subsection{Asymptotic behavior of the spectator functions}

The spectator function Eq.\eqref{regularspec} for shallow bound states ($\kappa_0 \rightarrow 0$) reads
 \begin{eqnarray}
&&\chi^{(i)}(q'_i)\underset{q_i  \gg \kappa_0}{=} \mathcal{C}^{(i)}
\kappa_{0}^{1-D} \mathfrak{F}_{(D,s_n)}\left(\frac{q_i^{\prime }}{\sqrt{2}\kappa_0}\right)^{1-D}\nonumber \\
&&\times\left[ \mathcal{G}_{(+ s_n)} \left(\frac{q_i^{\prime }}{\sqrt{2}\kappa_0}\right)^{s_n} +   \mathcal{G}_{(- s_n)} \left(\frac{q_i^{\prime }}{\sqrt{2}\kappa_0}\right)^{-s_n} \right],\ \ \ \
 \label{asympespec}
\end{eqnarray} 
where
 \begin{eqnarray}
 \mathcal{G}_{(\pm s_n)}  
=\frac{\Gamma(\pm s_n)}{\Gamma\left[(D-1\pm s_n)/2\right]\Gamma\left[(1\pm s_n)/2\right]},
 \label{eq:G}
\end{eqnarray} 
and $\mathfrak{F}_{(D,s_n)}$ is given by Eq.~\eqref{DefF}.
In the Efimov regime ($s_n \rightarrow i s_0$), the spectator function can be written as
\begin{eqnarray}
 \chi^{(i)}(q'_i) &\underset{q_i  \gg \kappa_0}{=}& \mathcal{C}^{(i)}\mathfrak{F}_{(D,is_0)}
 2\sqrt{\operatorname{Re}[\mathcal{G}_{(+is_0)}]^2+\operatorname{Im}[\mathcal{G}_{(+is0)}]^2} \nonumber \\
& \times&\left(\frac{q'_i}{\sqrt{2}} \right)^{1-D}\cos\left[ s_0 \ln \left(\frac{q'_i}{\sqrt{2}\kappa_0^{*}} \right)\right]\,,\ \ \ 
 \label{asympspecefimov}
\end{eqnarray}
where 
\begin{equation}    
\kappa_0^{*}\equiv \kappa_0/\exp\left\{\frac{1}{s_0}\arctan \left[\frac{\operatorname{Im}\ \mathcal{G}_{(+is0)}}{\operatorname{Re}\ \mathcal{G}_{(+is0)}}\right]\right\}.
\end{equation}
As expected, in this case the log-periodicity is present.

Further squeezing the system, one reaches the unatomic regime ($2<D<D_c$), where the values of $s_n$ are real and smaller than $0$. In this case, the spectator function becomes 
 \begin{eqnarray}
 \chi^{(i)}(q'_i)&\underset{q_i  \gg \kappa_0}{=} &\mathcal{C}^{(i)}\kappa_{0}^{1-D}
\mathfrak{F}_{(D,s_1)}\mathcal{G}_{(-s_1)} \left(\frac{q_i^{\prime }}{\sqrt{2}\kappa_0}\right)^{1-D-s_1}, \nonumber \\
 \label{asympespectunatomic}
\end{eqnarray} 
where the presence of the power-law symmetry, which is the signature of unatomic states, can be appreciated as
 \begin{eqnarray}
 \chi^{(i)}(\lambda\ q'_i)&\rightarrow&\lambda^{1-D-s_1}\chi^{(i)}( q'_i).
 \label{asympspectpowerlaw}
\end{eqnarray}

\begin{figure*}[!ht] 
{\includegraphics[width=8cm]{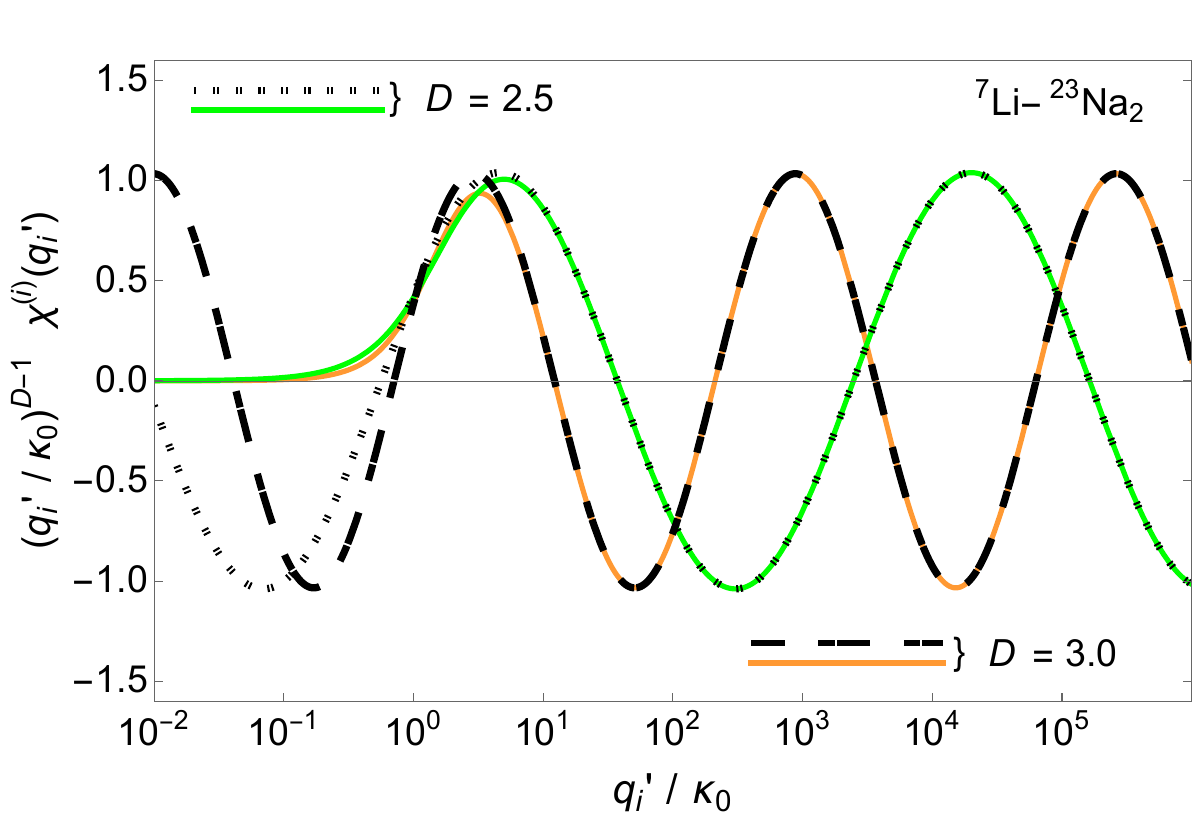}} 
{\includegraphics[width=8cm]{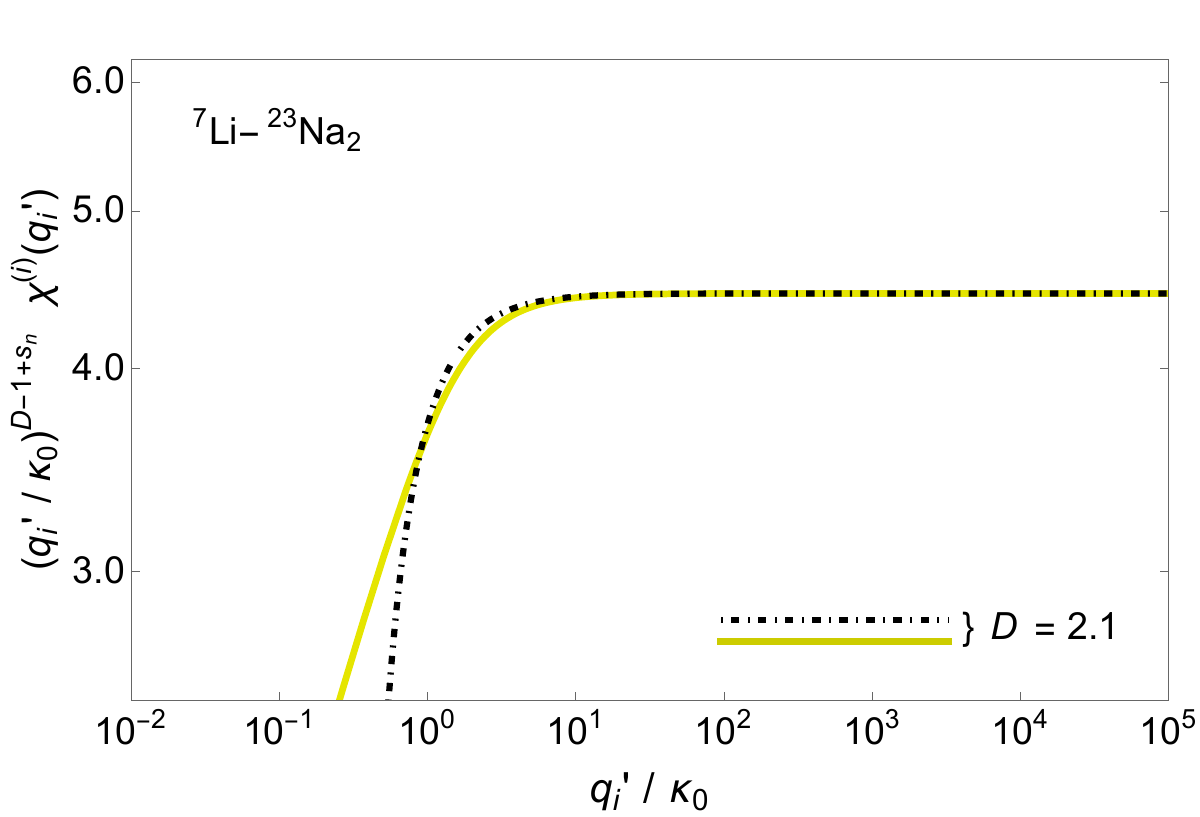}}
{\includegraphics[width=8cm]{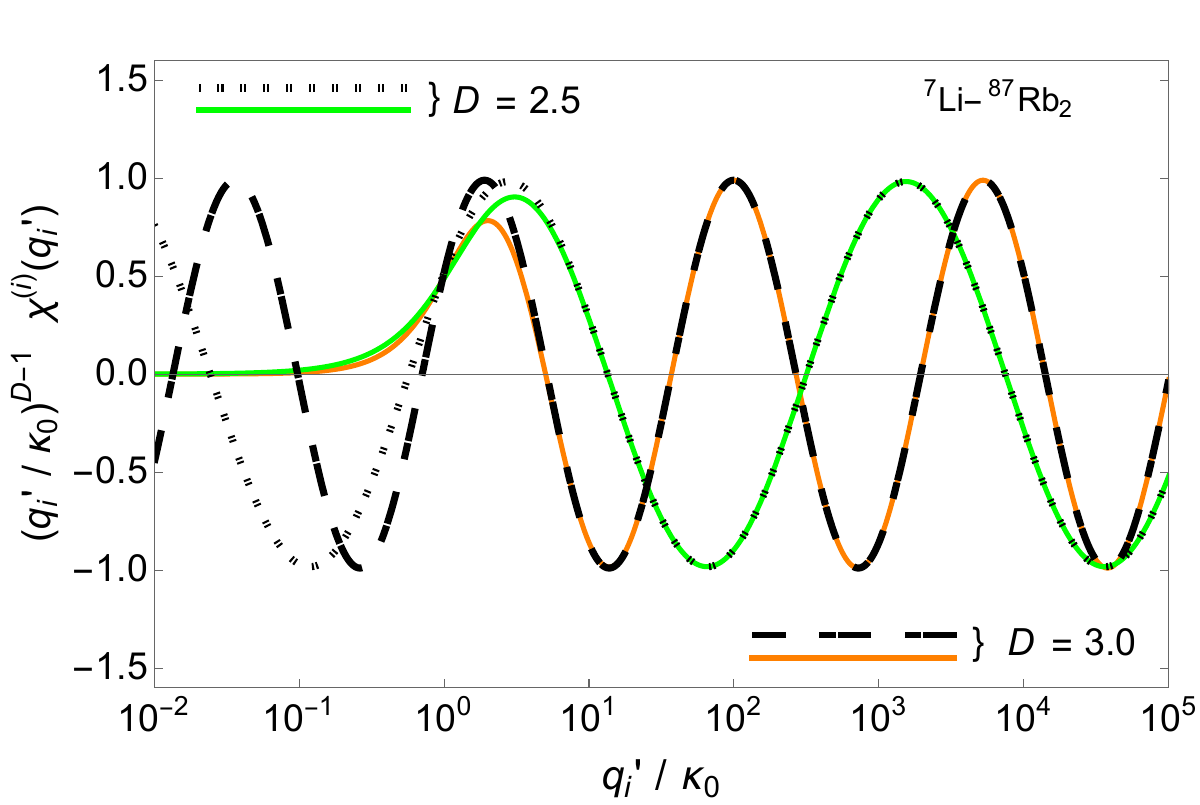}}
{\includegraphics[width=8cm]{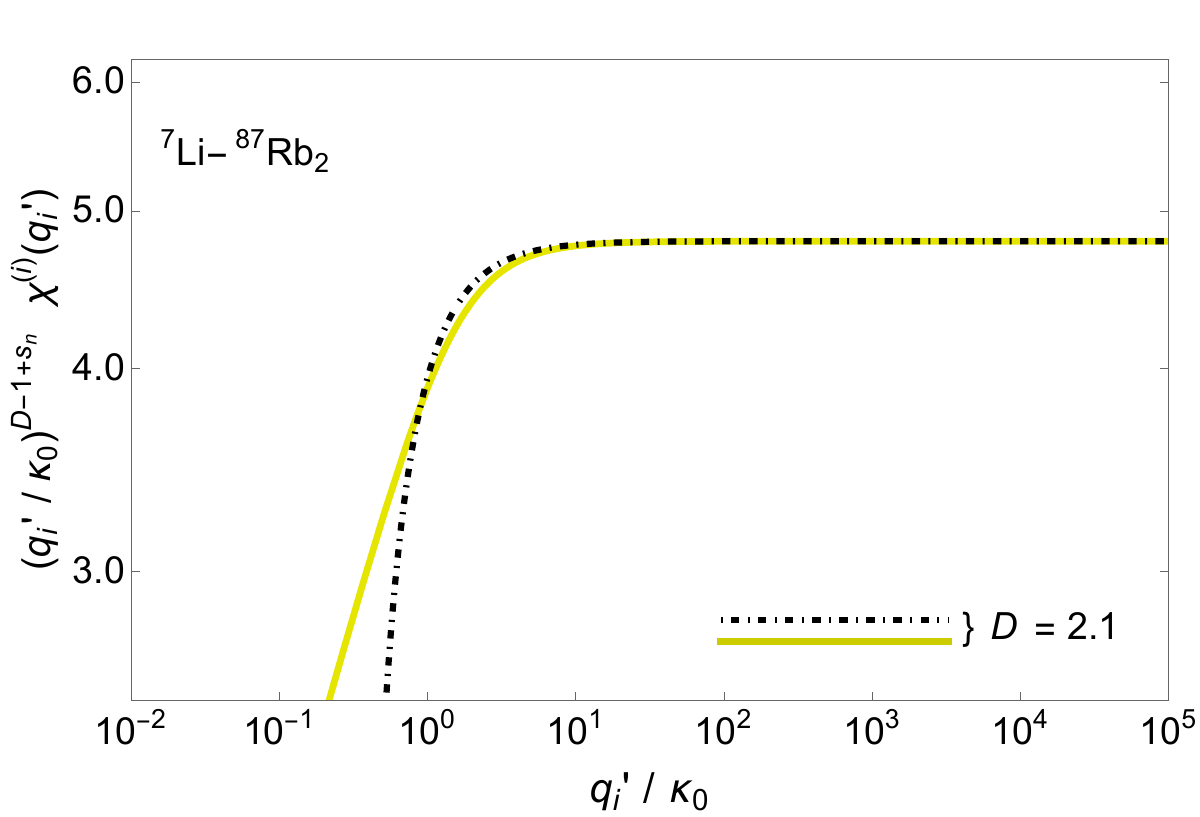}}
{\includegraphics[width=8cm]{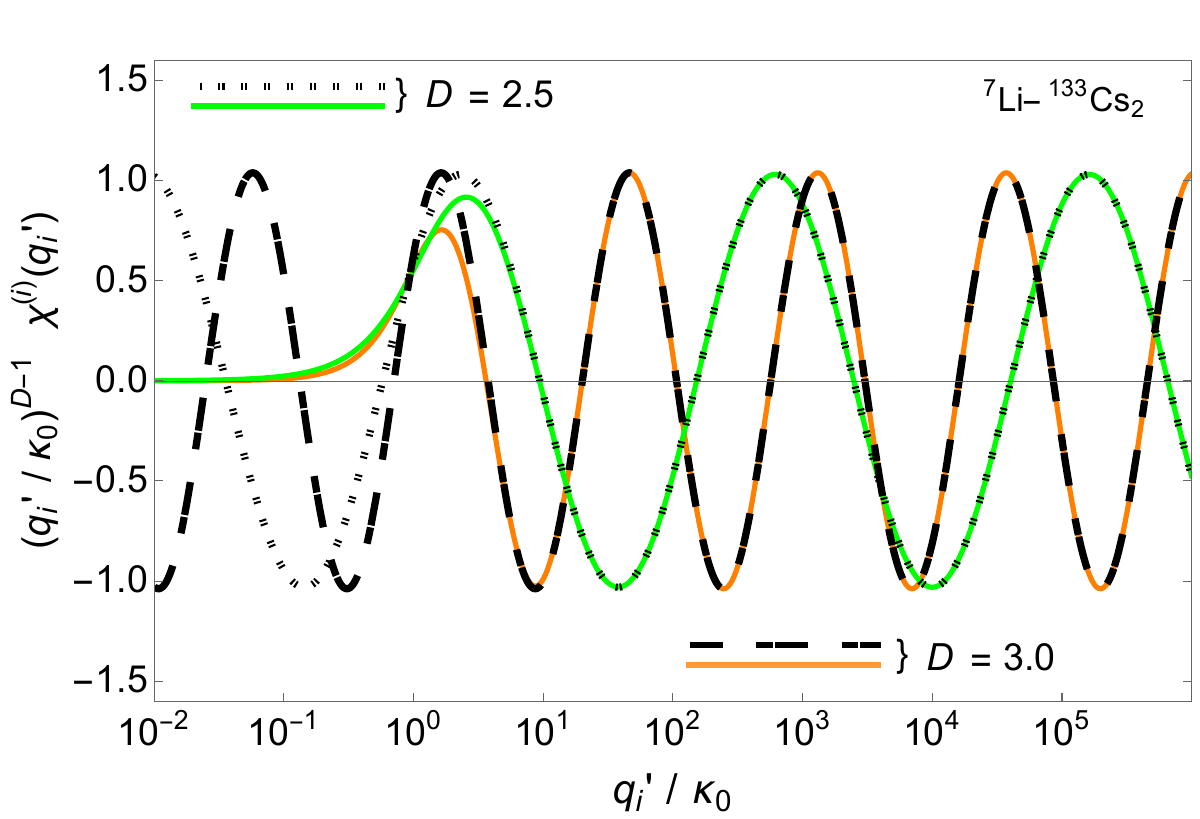}}
{\includegraphics[width=8cm]{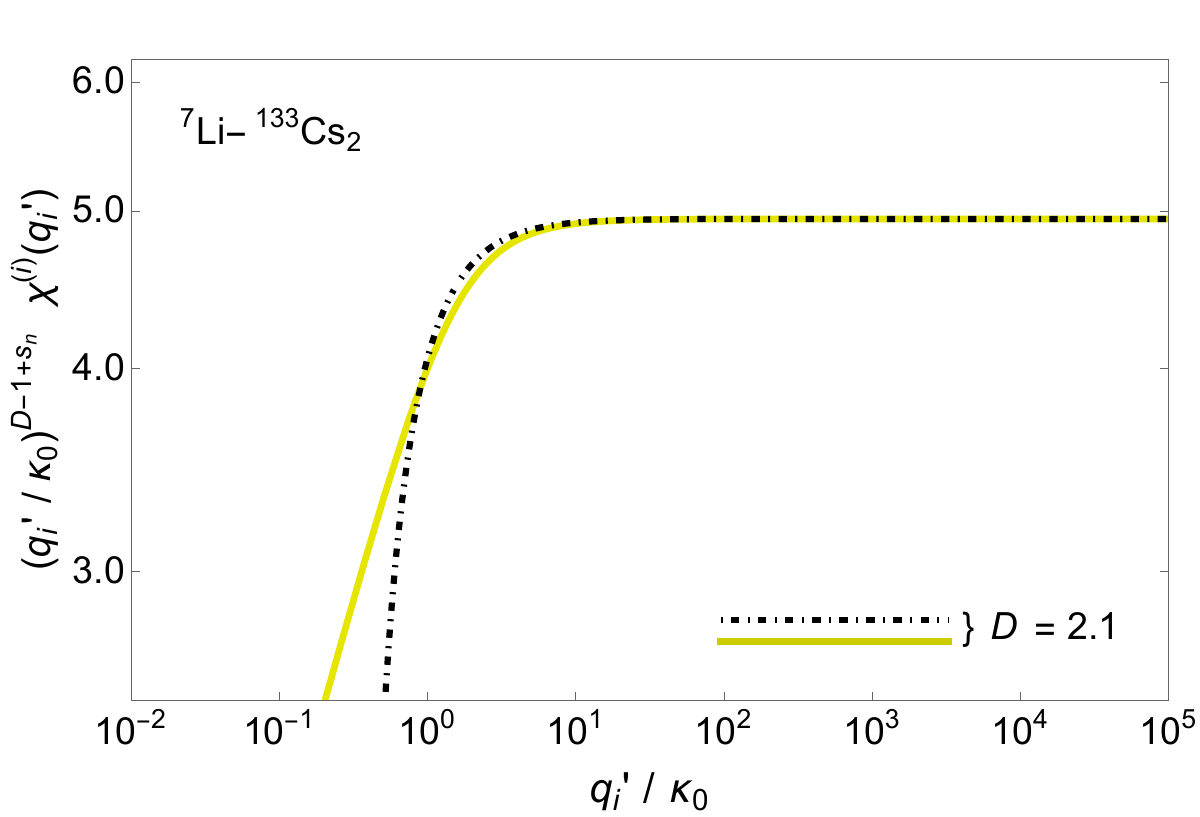}}
\caption{Spectator functions in momentum space for the $^{7}$Li$-^{23}$Na$_{2}$, $^{7}$Li$-^{87}$Rb$_{2}$ and $^{7}$Li$-^{133}$Cs$_{2}$ compounds. Left panels: $D=3$ and $D=2.5$. Finite three-body energy (colored lines) and zero-energy (long-dashed and dashed black lines). Right panels: $D=2.1$. Finite three-body energy (colored lines) and zero-energy dotted-dashed}.
\label{fig2}
\end{figure*}

In Fig.~\ref{fig2}, we contrast the finite three-body energy spectator function~\ref{regularspec} with its asymptotic behavior~\ref{asympespec} for the $^{7}$Li-AA systems embedded in different dimensions. In the left panels, the results for the Efimov regime are exhibited. As already broadly discussed, the decrease in the effective dimension impacts the three-body system by decreasing the scale parameter $s_0$, so that the distance between consecutive nodes of the wave-function increases. Besides that, an increase in the  mass imbalance of the system makes the scale parameter bigger, so that the distance between consecutive nodes of the wave function decreases. In the right panels, the spectator functions are displayed for the unatomic regime, where the power-law fingerprint of the unatomic regime starts to be manifested. In both left and right panels, the spectator functions obtained for finite three-body binding energy (colored lines) are characterized by a damping in the low momentum region.

\subsection{Momentum distribution}
 \label{section3}
 
The momentum distributions in $D$-dimensions for the particles $A$ and $B$ are given, respectively, by
\begin{equation}
 n_A(q_A) = \int d^{D}p_A \ |\langle \textbf{q}_A \textbf{p}_A | \Psi \rangle |^{2} \,,
 \end{equation}
 and
\begin{equation}
 n_B(q_B) = \int d^{D}p_B \ |\langle \textbf{q}_B \textbf{p}_B | \Psi \rangle |^{2} \,,
 \label{momdis}
 \end{equation}
\noindent The normalization conditions are taken to be
\begin{equation} 
 %\int d^Dr\ d^D\rho \ |\Psi(\textbf{r}, \pmb{\rho} )|^2
 \int d^Dq_B\, n_B(q_B)
 =1\,\,\text{and}\, \,
 \int d^Dq_A\, n_A(q_A)
 =1
 \, .
 \label{norm}
 \end{equation}  

We illustrate our study focusing on the momentum distribution of the light particle. By means of the Faddeev decomposition shown in Eq.~\ref{A11}, the momentum density in Eq.~\ref{momdis} can be split
into nine terms, which, because of the symmetry between the two identical particles $A$, are reduced to  
 \begin{equation}
 \label{4sum}
 n_B(q_B) = n_1(q_B) + n_2(q_B) + n_3(q_B) + n_4(q_B),
 \end{equation}
where

 \begin{eqnarray}
n_{1}(q_B)= \lvert \chi^{(B)}(q_B) \rvert^{2} \int d^{D}p_B \frac{1}{\left(E_3 +  p_{B}^{2}+q_{B}^{2}/2\mu_B\right)^{2}},\ \ \ \ \   \nonumber \\
 \label{n1}
 \end{eqnarray}
 \begin{eqnarray}
 n_{2}(q_B) = 2 \int d^{D}p_B \frac{\lvert \chi^{(A)}(\lvert \textbf{p}_B -\textbf{q}_B/2 \rvert) \rvert^{2}}
 { \left( E_3 + p_{B}^{2}+ q_{B}^{2}/2\mu_B \right)^{2} }, \hspace{2.1cm}
 \label{n2}
\end{eqnarray}
 \begin{eqnarray}
 n_{3}(q_B) &=& 2  [\chi^{(B)}(q_B) ]^*\int d^{D}p_B \frac{\chi^{(A)}(\lvert \textbf{p}_B - \textbf{q}_B/2  \rvert )  }
 { \left( E_3 + p_{B}^{2} + q_{B}^{2}  /2\mu_B \right)^{2} },\ \ \ \ \ \ \nonumber \\
 &+& {\rm c.c.},
 \label{n3}
 \end{eqnarray}
 \begin{eqnarray}
 n_{4}(q_B) &=& \int d^{D}p_B \frac{[\chi^{(A)}\lvert \textbf{p}_B-\textbf{q}_B/2 \rvert)]^*
 \chi^{(A)}(\lvert \textbf{p}_B + \textbf{q}_B/2  \rvert )  }
 {\left( E_3 + p_{B}^{2} + q_{B}^{2}  /2\mu_B\right)^{2}} \nonumber \\
 &+& {\rm c.c.}.
 \label{n4}
 \end{eqnarray} 
The computation of Eqs.~\eqref{n1} to~\eqref{n4} with the spectator function written in Eq.\eqref{regularspec} allows us to calculate the regular momentum distribution $n_B(q_B)$ for different noninteger dimensions. 

\begin{center}
\begin{figure}[h!] 
{\includegraphics[width=8cm]{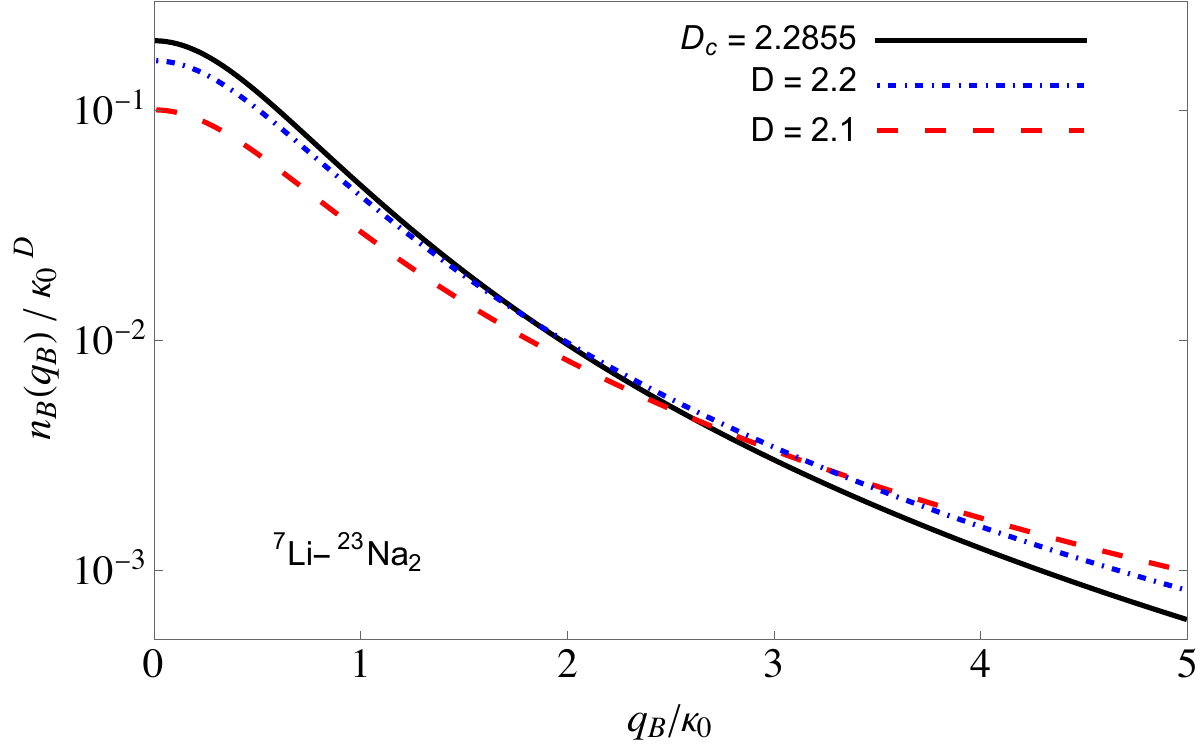}}
{\includegraphics[width=8cm]{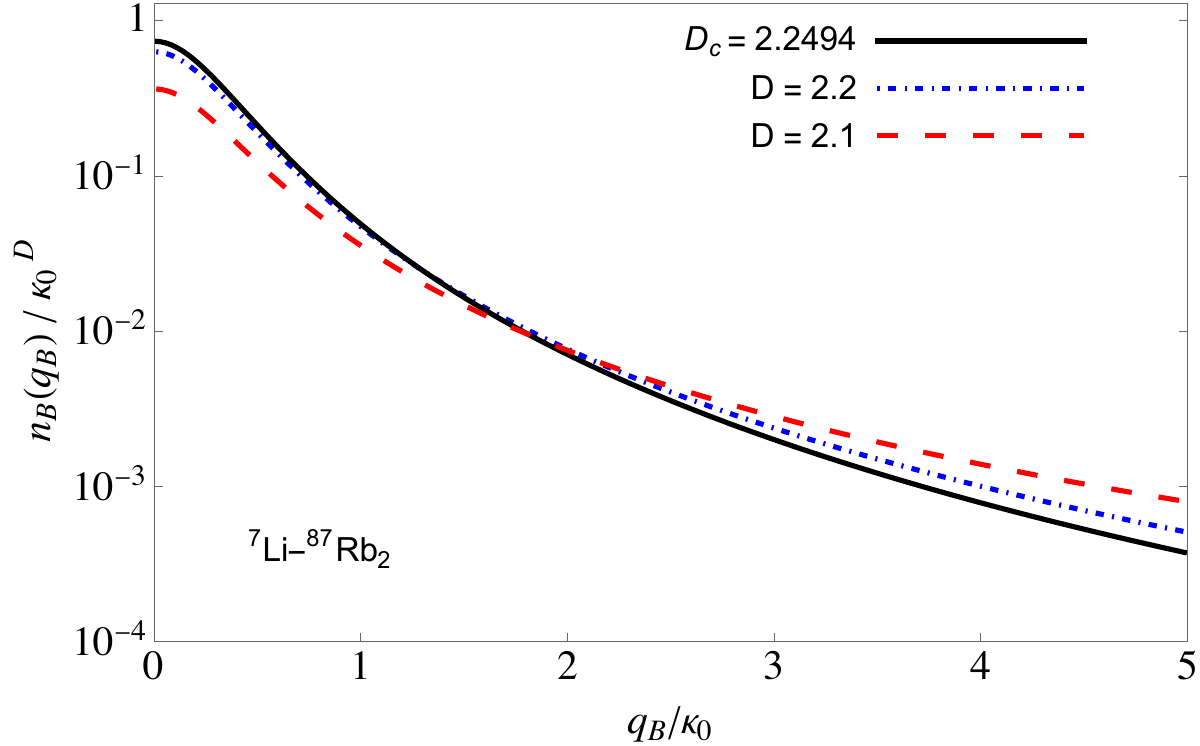}}
{\includegraphics[width=8cm]{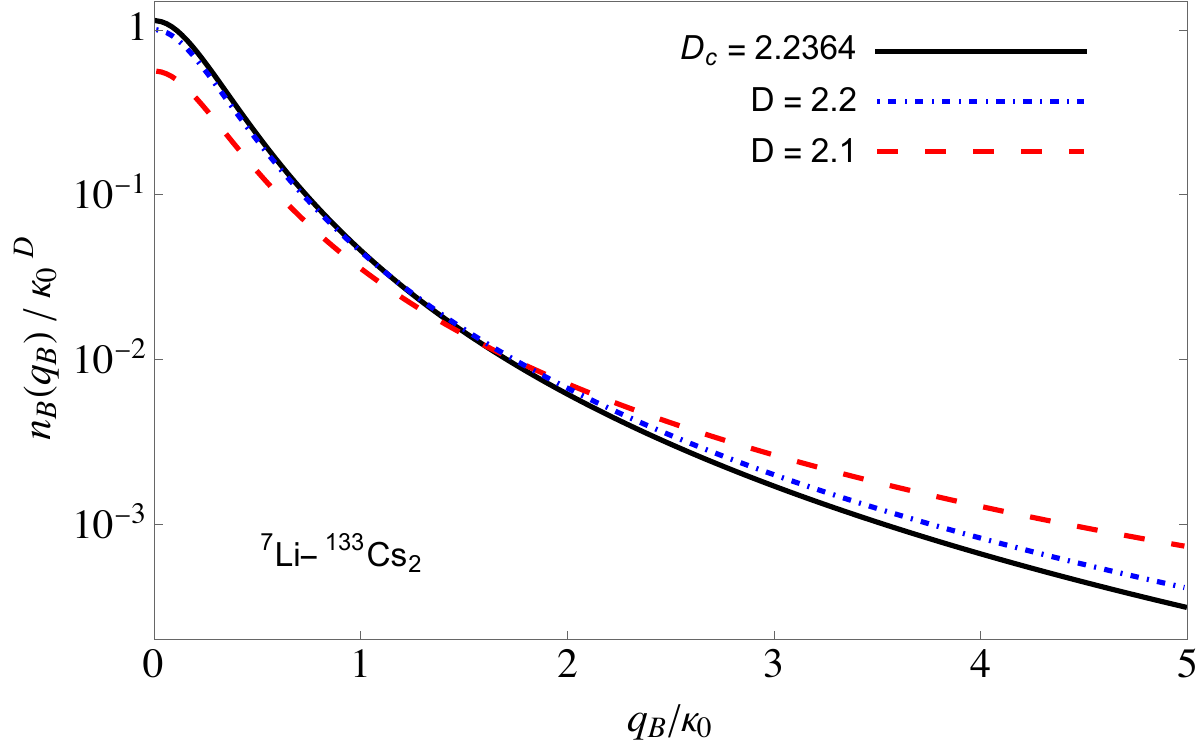}}
\caption{Single particle momentum distribution $ n_B(q_B)$ of the compounds $^{7}$Li$-^{23}$Na$_{2}$, $^{7}$Li$-^{87}$Rb$_{2}$ and $^{7}$Li$-^{133}$Cs$_{2}$ for different dimensions.}
\label{fig3}
\end{figure}
\end{center}

The normalized momentum density, $n_B(q_B)$, is displayed in Fig.~\ref{fig3} for the $^{7}$Li-AA systems in the noninteger dimensions $D_c$, $D=2.2$ and $D=2.1$. We observe that the squeezing of the system, by lowering the noninteger dimension, tends to emphasize the large momentum region~\cite{Nb_AAB_Ddim_Efimov}, so that the system becomes less localized. This effect goes until $D=2$, where the system reaches the continuum. As we will show, this enhancement of the large momentum region is also expressed in the two- and three-body contacts. Besides that, we observe that lowering the mass imbalance of the system makes the amplitude of the momentum distributions smaller. 

We now discuss the leading and sub-leading contributions of the large-momentum tail of the single-particle momentum distribution from which one can extract the contact parameters, quantities that can provide a link between theoretical and experimental realizations. Detailed steps for the calculations of the sub-leading contributions in the Efimov and unatomic regimes can be found in appendices~\ref{appb} and~\ref{appd}, respectively.

To start, we have to insert asymptotic spectator functions for each spatial region of interest in Eqs.~\eqref{n1}-\eqref{n4} and perform the corresponding integrals. The contribution $n_1(q_B)$, Eq.\eqref{n1}, is straightforward to calculate
 \begin{eqnarray}
 n_{1}(q_B) &=&  \frac{\lvert \chi^{(B)}(q_B) \rvert^{2}}{q_{B}^{4-D}} \mathcal{S}_{D}    \frac{\pi}{4}\csc\left( \frac{D\pi}{2} \right) (2-D)\nonumber \\ &\times&\left(2\mu_B\right)^{2-D/2} \, ,
 \label{eq:n1desenvolved}
 \end{eqnarray} 
where $\mathcal{S}_{D}$ is the area of a $D$-dimensional sphere. 

The second contribution, $n_2(q_B)$, can be computed 
from Eq.~\eqref{n2} by making the change of variables $\textbf{p}_B-\textbf{q}_B/2=\textbf{q}_A$, that results in 
\begin{equation}
 n_{2}(q_B) = 2 \int d^{D}q_A \frac{\lvert \chi^{(A)}(q_A) \rvert^{2}}
 { \left(  q_A^{2}+ \textbf{q}_A.\textbf{q}_B + q_{B}^{2}  \mu_A/\mu_B \right)^{2} }\, .
 \end{equation}
In order to identify the leading order term, we perform the 
manipulation~\cite{castindensity,yamashita2013}
\begin{eqnarray}\label{eq:n2}
 n_{2}(q_B) &=& 2 \int d^{D}q_A\lvert \chi^{(A)}(q_A) \rvert^{2}\nonumber \\
 &\times&\left[ \frac{1}
 { \left( q_A^{2}+ \textbf{q}_A.\textbf{q}_B + q_{B}^{2}  \mu_A/\mu_B \right)^{2} }-\left(\frac{\mu_B}{\mu_A }\right)^2\frac{1}{q_B^4}\right] \nonumber \\
 & +& \frac{C_2}{q_B^{4}}\, ,
 \end{eqnarray}
where $C_2$ is the two-body contact, which can be related to the derivative w.r.t. the scattering length of the gas's mean energy (or  mean free energy
at nonzero temperature)~\cite{castin2d3d,WernerPRA2008}. It has dimension  
$(\text{length})^{D-4}$ and therefore  scales as $C_2\propto \kappa_0^{4-D}$. The two-body contact parameter is given by

 \begin{eqnarray}
 C_{2}
= 2\left(\frac{\mu_B}{\mu_A }\right)^2\mathcal{S}_D \int^\infty_0 dq_A \, q_A^{D-1}\lvert \chi^{(A)}(q_A) \rvert^{2}\,. \label{eq:c2}
 \end{eqnarray}
In spherical coordinates, Eq.~\eqref{eq:n2} reads
\begin{eqnarray}
 &&n_{2}(q_B) = \frac{2(2\pi)}{q_{B}^{4}}\prod_{k=1}^{D-3}\int_0^{\pi} d\theta_k\sin^{k}\theta_k \nonumber \\
 &&\times\int^\infty_0 dq_A\   q_A^{D-1}  | \chi^{(A)}(q_A) |^{2}
\int_0^{\pi} d\theta\ \sin^{D-2}\theta \nonumber \\
&&\times\left\{\frac{1}{\left[ (q_A/q_B)^{2} + (q_A/q_B)\cos\theta+ \mu_B/\mu_A \right]^{2}}\right. \nonumber \\
&&\left.- \left(\frac{\mu_B}{\mu_A}\right)^2\right\}+\frac{C_2}{q_B^4}\,.
 \end{eqnarray}
\noindent Making the change of variables $q'_A=q_A/ q_B$, one finds
\begin{eqnarray}
n_{2}(q_B) &= & \frac{2\mathcal{S}_D }{q_B^{4-D}}\int^\infty_0 dq'_A \, q_A^{\prime D-1} | \chi^{(A)}(q_B\, q'_A) |^{2}\nonumber \\ &\times& \left(\mathcal{H}(q'_A)-\frac{4 \mathcal{A}^2}{(\mathcal{A}+1)^2}\right)+\frac{C_2}{q_B^{4}},
 \label{eq:n2desenvolved}
 \end{eqnarray}
 with
  \begin{eqnarray}
 \mathcal{H}(y)&\equiv&\frac{4 \mathcal{A}^2 (D-2)}{\mathcal{A}^2 \left(4 y^4+1\right)+\mathcal{A} \left(4 y^2+2\right)+1} \nonumber \\
 &+&\frac{4 \mathcal{A}^2 (3-D) \left(2  y^2\ \mathcal{A}+\mathcal{A}+1\right) \, }{[2 \mathcal{A} (y-1) y+\mathcal{A}+1]^2 [2  y (y+1)\mathcal{A}+\mathcal{A}+1]}\nonumber \\
 &\times& H_2F_1\left(1,\frac{D-1}{2},D-1,-\frac{4 \mathcal{A} y}{2 (y-1) y \mathcal{A}+\mathcal{A}+1}\right),\nonumber \\
 \label{eq:H2F1}
 \end{eqnarray}
where $H_2 F_1(a,b,c,z)$ is the hyper-geometrical function and $\mathcal{A}\equiv m_B/m_A$. 

In the large momentum limit ($q_B\gg \sqrt{2\mu_B E_3}$), after the change of variables $\textbf{p}_B - \textbf{q}_B/2 = \textbf{q}_A$ and the observation that the spectator functions are real, the third contribution, $n_3(q_B$), given in Eq.~\eqref{n3}, can be written as

  \begin{equation}
n_{3}(q_B)=  4\int d^{D}q_A \frac{[\chi^{(B)}(q_B)]^* \chi^{(A)}(q_A )  }
 { \left( q_{A}^{2} +\textbf{q}_A.\textbf{q}_B +q_B^{2}\mu_A/\mu_B \right)^{2} }\,.
 \end{equation}
Changing variables as $q_A/q_B =q'_A$ and integrating in spherical coordinates, one obtains:
\begin{equation}
n_{3}(q_B)= [\chi^{(B)}(q_B)]^*\frac{4\mathcal{S}_D}{q_B^{4-D}} 
\int^\infty_0 \hspace{-.3cm}  dq'_A  \ q_A^{\prime D-1}  \chi^{(A)}(q_B q'_A)\mathcal{H}(q'_A)\,,
\label{eq:n3desenvolved}
\end{equation}
where $\mathcal{H}(y)$ is  given by Eq.~\eqref{eq:H2F1}. 

For the contribution $n_4(q_B)$, we observe that there is an argument in the spectator function that can be written as
\begin{eqnarray}
 \lvert \textbf{p}_B\pm\frac{\textbf{q}_B}{2}\rvert = q_B \sqrt{\frac{p_B^{2}}{q_B^{2}}+\frac{1}{4}\pm \frac{p_B}{q_B}\cos\theta},
\end{eqnarray}
where, making the same steps as in the previous contribution, but now changing the variables as $p_B /q_B =  p'_B$, one find
 \begin{eqnarray}
 n_{4}(q_B) &= & \frac{1}{q_B^{4-D}} 4\pi\prod_{k=1}^{D-3}\int_0^{\pi}d\theta_k  \sin^{k}(\theta_k)\nonumber   \\
 &\times&\int^\infty_0 dp'_B  \frac{p_B^{\prime D-1}  }
 { \left( p_{B}^{\prime 2} + 1/2\mu_B \right)^{2} } \int_0^{\pi}d\theta \sin^{D-2}\theta \nonumber  \\
 &\times& [\chi^{(A)}( q_B \, p'_{B -})]^*\chi^{(A)}(q_B \, p'_{B +})\, ,
 \label{eq:n4desenvolved}
 \end{eqnarray}
 where $ p'_{B\pm}=\sqrt{p^{\prime 2}_B+\frac14\pm p'_B\cos\theta}$.
 
In the Efimov regime, from the integral representations in  Eqs.~\eqref{eq:n1desenvolved},~\eqref{eq:n2desenvolved},~\eqref{eq:n3desenvolved} and \eqref{eq:n4desenvolved}, one can obtain the oscillatory and non-oscillatory contributions to each of the four contributions of the momentum density at large momentum by using the asymptotic spectator function given by Eq.~\eqref{asympspecefimov}. The leading and sub-leading contributions in the asymptotic region are given by
\begin{eqnarray}
n_B(q_B)&\underset{q_B  \gg \kappa_0}{=}&  \frac{C_{2}}{q_{B}^{4}}  + \frac{C^{'}_{3}}{q_{B}^{D+2}} 
\nonumber \\
& +&  \frac{C_{3}}{q_{B}^{D+2}} \cos\!\left[2 s_0 \ln\left(\frac{q_B/\kappa_0^{*}}
{(4\mu_A \mu_B )^{1/4} }  \right) \! + \!\Phi \right] \nonumber \\
&+& \cdots, \ \ \ \ \ \ ~\text{for}~  D_c <D \leq 3, \ \ \ 
\label{c3c3l}
\end{eqnarray} 
where the amplitudes $C_3^{'}$, $C_2$ and $C_3$ describe the non-oscillatory behavior, while $\Phi$ represents the phase related to the log-periodic oscillatory term. The parameter $C_3$ is the three-body contact, closely related to the Efimov effect as it gives the amplitude of the log-periodic function of the momentum distribution~\cite{braaten,braaten2014}.

Below the critical dimension, the high momentum regime of the single particle momentum distribution is computed with the asymptotic spectator function that presents a power law symmetry, Eq.~\eqref{asympespectunatomic}. A narrow dimensional region is defined below the critical dimension ($\overline D < D < D_c$), where the two-body parameter still contributes to the tail of the momentum distribution. Adding the results of the four contributions, we obtain:
\begin{eqnarray}
n_B(q_B) &\underset{q_B  \gg \kappa_0}{=} & \frac{C_{2}}{q_B^{4}}  +C_{3}\left[\frac{q_B/\kappa_0}
{(4\mu_A \mu_B )^{1/4} }  \right]^{-(D+2+2s_1)}\nonumber \\
&+& \cdots , \ \ \ \
~ \text{for}~   \overline D < D<D_c.
\label{nbc2ceunatomic}
\end{eqnarray}
In Eq. \ref{nbc2ceunatomic}, we see that the tail of the momentum distribution, although in the region of real scale parameter, does not display symmetry. This happens because the presence of the two-body parameter breaks such a behavior. For dimensions below $\overline{D}$, the tail of the momentum distribution is entirely defined by the three-body parameter, namely: 
\begin{eqnarray}
n_B(q_B) \underset{q_B  \gg \kappa_0}{=} \frac{{C}_{3}}{q_B^{D+2+2s_1}} + \cdots, ~ \text{for}~  2< D\leq \overline D.\ \ \ \ 
\label{nbunatomic}
\end{eqnarray} 
In this regime, the single-particle momentum distribution tail exhibits a power-law behavior, $1/q^{D+2+2s_1}$. This power-law behavior characterizes the unatomic nature of the system, in that it displays the scaling law
\begin{equation}
n(\lambda\ q) \rightarrow \lambda^{\Delta_{n}} n( q) \,,
\end{equation}
where $\Delta_{n} \equiv -(D+2+2s_1)$.

\begin{center}
\begin{figure}[h!]
\includegraphics[width=8.5cm]{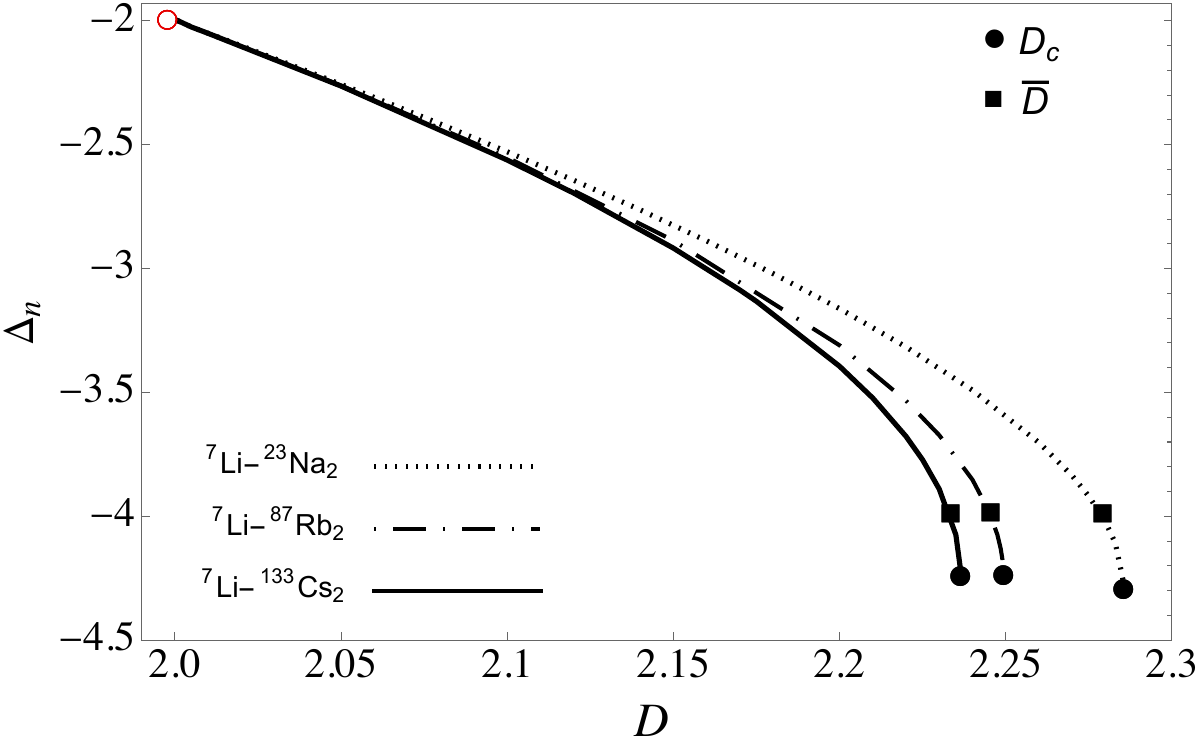}
\caption{Scaling coefficient parameters for the compounds of the form $^7$Li-AA. The black dot identify the critical dimension, given by $D_c = 2.2855$, $2.2494$ and $2.2364$ for the $^{7}$Li$-^{23}$Na$_{2}$, $^{7}$Li$-^{87}$Rb$_{2}$, and $^{7}$Li$-^{133}$Cs$_{2}$ compounds, respectively. The square identify the dimension at which there occurs an exchange between the leading and sub-leading order terms in the asymptotic expansion. These dimensions are given by $\overline{D} = 2.2792$, $2.2455$, and $2.2334$, for the compounds $^7$Li-$^{23}$Na$_2$, $^7$Li-$^{87}$Rb$_2$ and $^7$Li-$^{133}$Cs$_2$, respectively. The empty red circle indicates the point where the resonant three-atom system reaches the threshold.}
\label{fig4}
\end{figure}
\end{center}

Fig.~\ref{fig4} shows the scaling coefficient for the compounds of the form $^7$Li-AA embedded in noninteger dimensions below the critical one. The black points denote the critical dimensions where the transition from discrete to continuum scale symmetry occurs. At the dimensions $\overline{D}$, there is an exchange between the leading and sub-leading order terms in the asymptotic expansion of the single particle momentum distribution, Eq.~\ref{nbc2ceunatomic}. At these points, indicated by black squares, we have $\Delta_{n}=-4$. Below $\overline{D}$, the unatomic state defines the properties of the observables. The empty red circle indicates where the resonant unatomic state reaches the threshold, being the three-body bound energy proportional to the two-body one at $D=2$. 

%%%%%%%%%%%%%%%%FIGURE04%%%%%%%%%%%%%%%%%%%
%%%%%%%%%%%%%%%%%%%%%%%%%%%%%%%%%%%%%%%%%%%%%

\subsection{Contact Parameters} \label{sec:contact} 

The contact parameters for the Efimov states are extracted from the log-periodic asymptotic density, Eq.~\eqref{c3c3l}. $C_3$ and the phase $\Phi$ are computed by adding Eqs.~\eqref{eq:A3}, \eqref{eq:B7},  \eqref{eq:C4} and \eqref{eq:D4}. $C'_3$ is obtained by adding  Eqs.~\eqref{eq:A4}, \eqref{eq:B8}, \eqref{eq:C5} and \eqref{eq:D5}. Both parameters $C_3$ and $C'_3$ scale with $\kappa^2_0$, or, equivalently, the three-body bound-state energy. $C_2$ is given by Eq.~\eqref{eq:c2} and is computed using the regular spectator function, Eq.~\eqref{regularspec}.

%%%%%%%%%%%%%%%%FIGURE05%%%%%%%%%%%%%%%%%%%
\begin{center}
\begin{figure}[h]
\includegraphics[width=8.7cm]{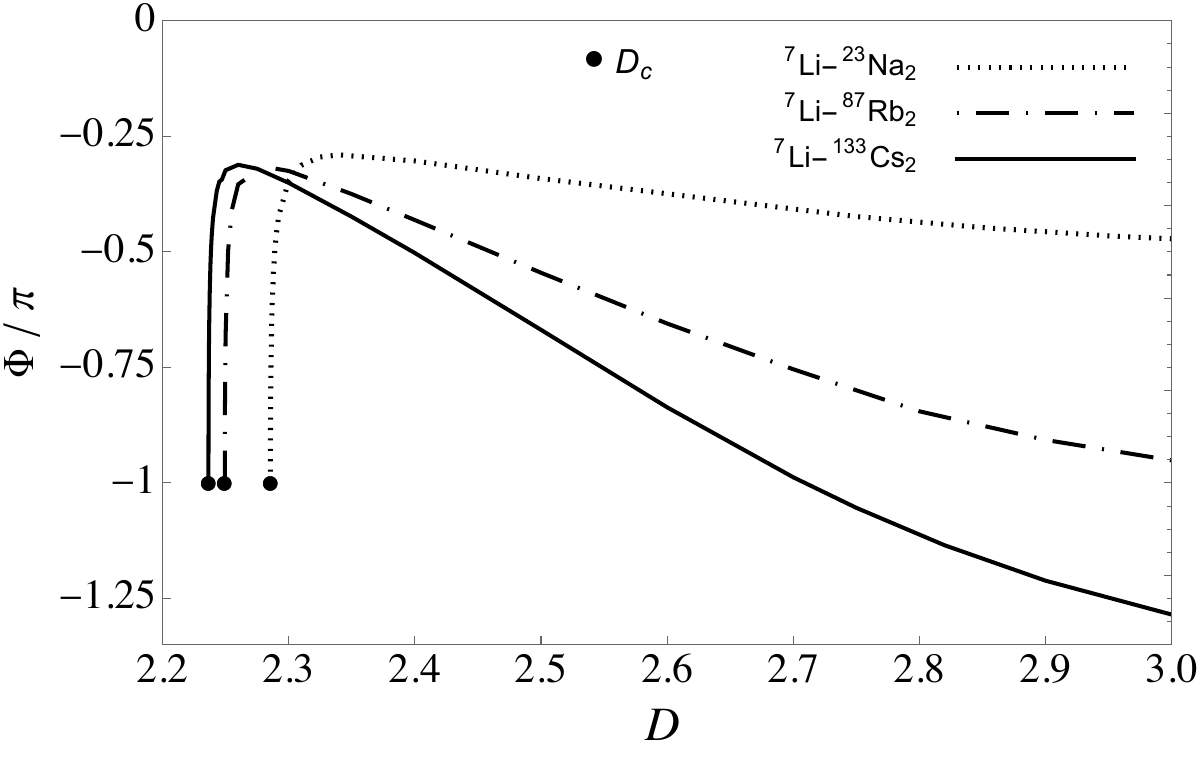}
\caption{
The phase factor~$\Phi$ as function of the noninteger dimension~$D$. The black dot identifies the critical dimension, which are given by $D_c = 2.2855$, $2.2494$ and $2.2364$ for the $^{7}$Li$-^{23}$Na$_{2}$, $^{7}$Li$-^{87}$Rb$_{2}$, and $^{7}$Li$-^{133}$Cs$_{2}$ compounds, respectively.}
\label{fig5}
\end{figure}
\end{center}
%%%%%%%%%%%%%%%%%%%%%%%%%%%%%%%%%%%%%%%%%%%%%

Figure~\ref{fig5} shows the $D$ dependence  of the phase $\Phi$. The phase increases as $D$ decreases, approaching sharply the critical point $\Phi=-\pi$. Furthermore, smaller mass imbalances result in a smoother curve, as evidenced by the comparison of the curves for the $^{7}$Li$-^{23}$Na$_{2}$ (dotted line), $^{7}$Li$-^{87}$Rb$_{2}$ (dashed-dotted line), and $^{7}$Li$-^{133}$Cs$_{2}$ (solid line) compounds. 

\begin{figure*}[t!]
\includegraphics[width=8.5cm]{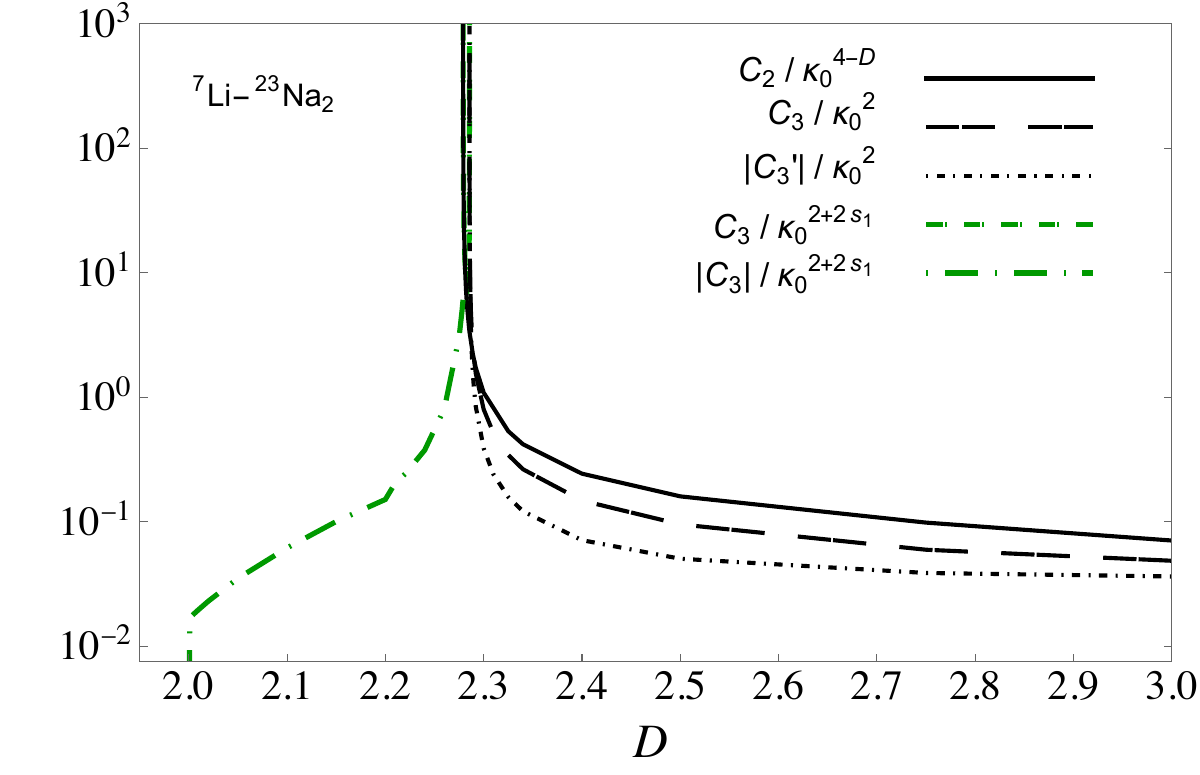}
\includegraphics[width=8.5cm]{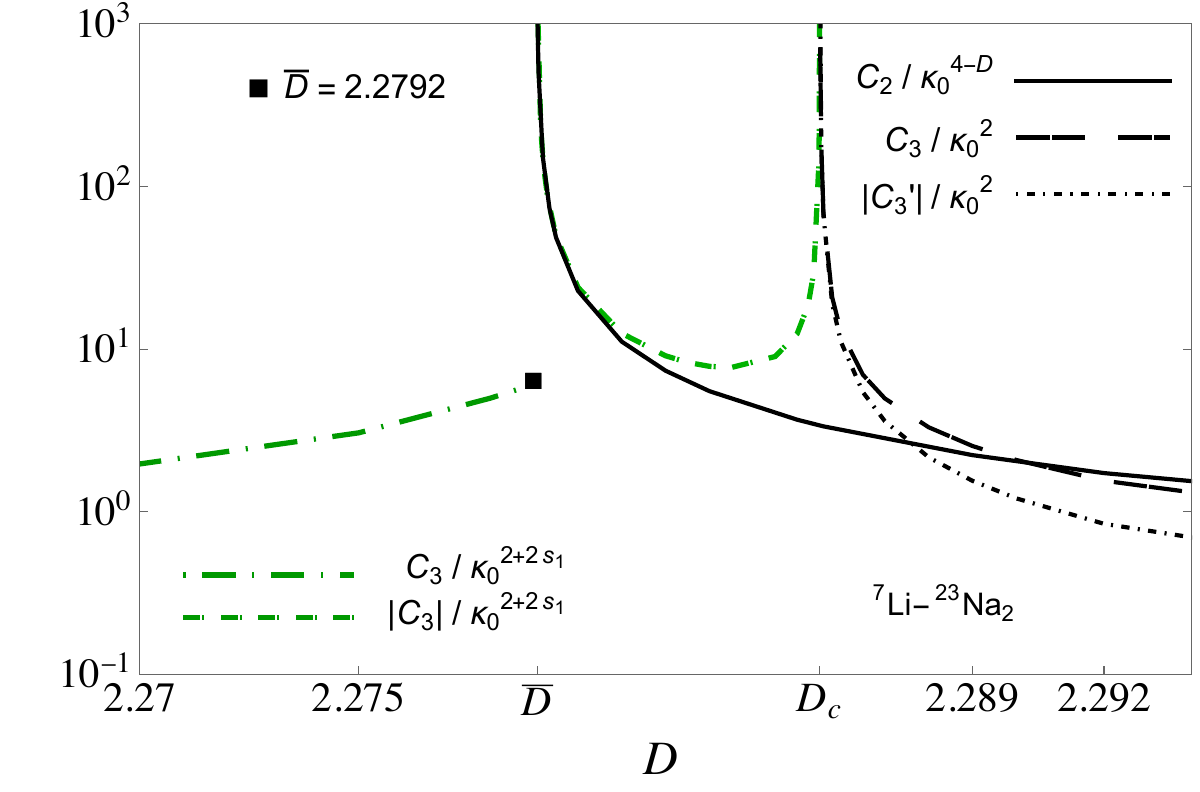}
\includegraphics[width=8.5cm]{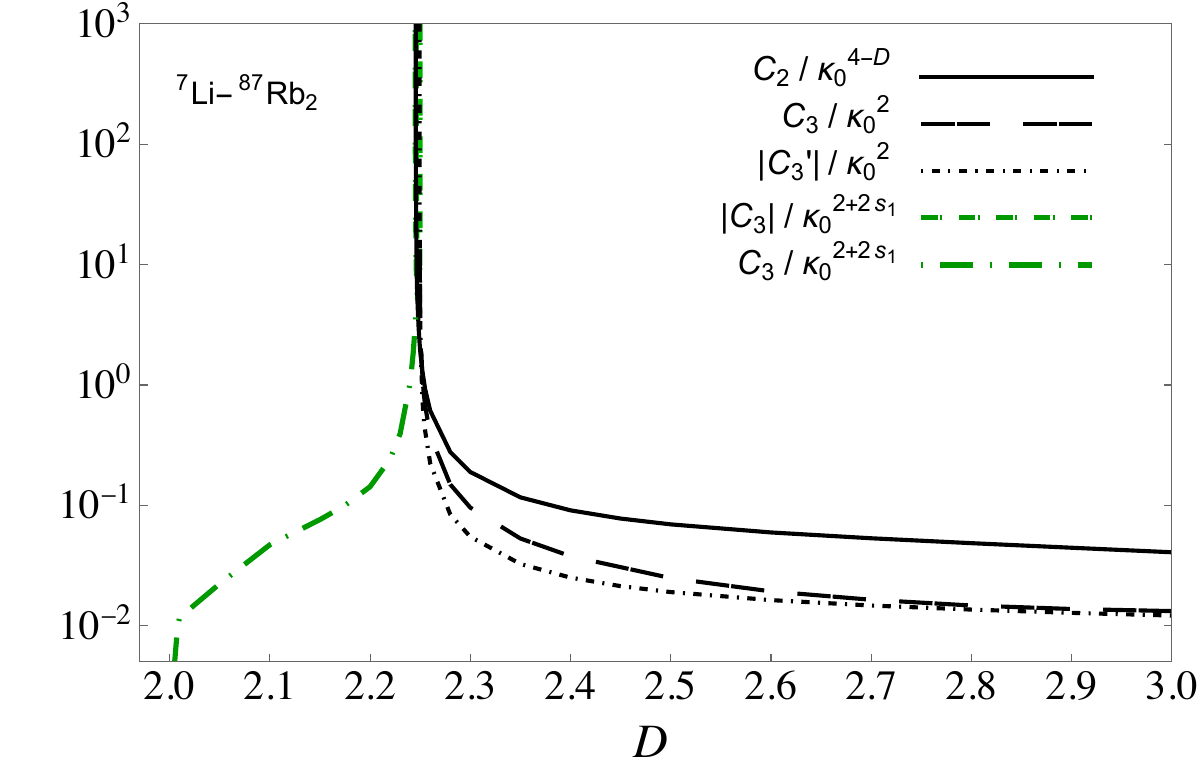}
\includegraphics[width=8.5cm]{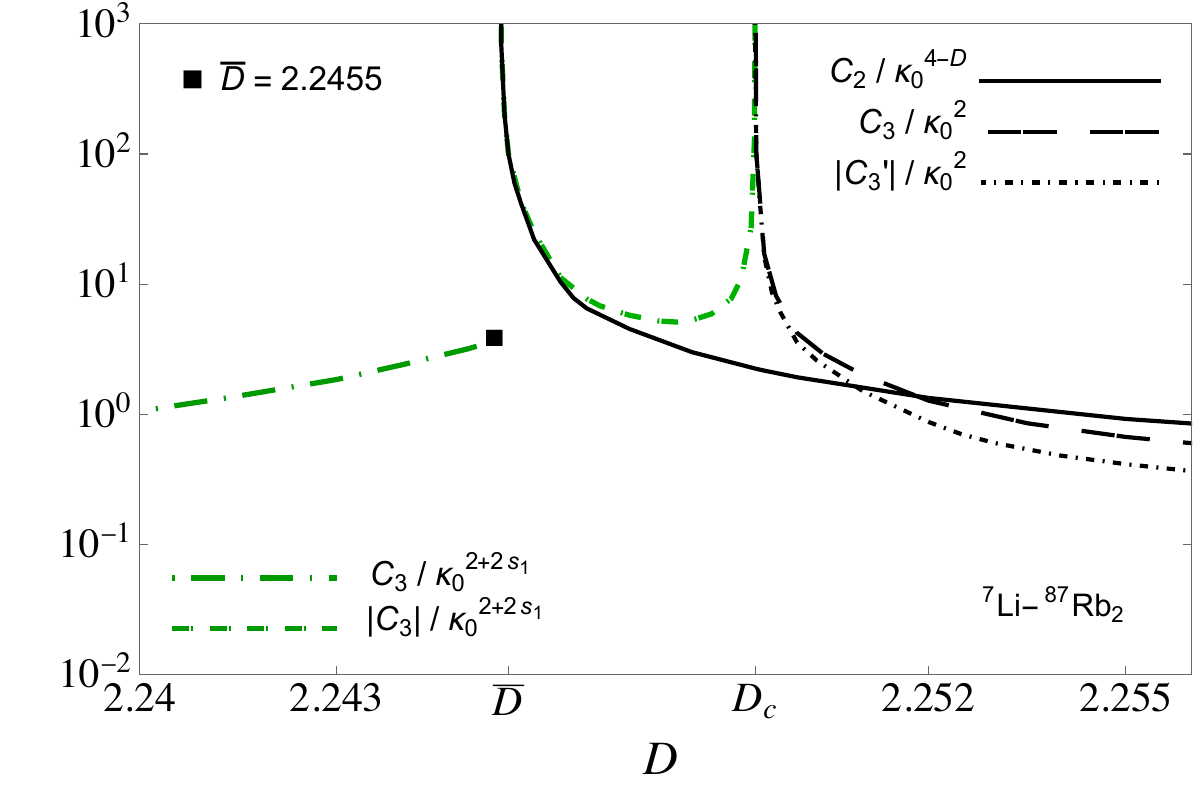}
\includegraphics[width=8.5cm]{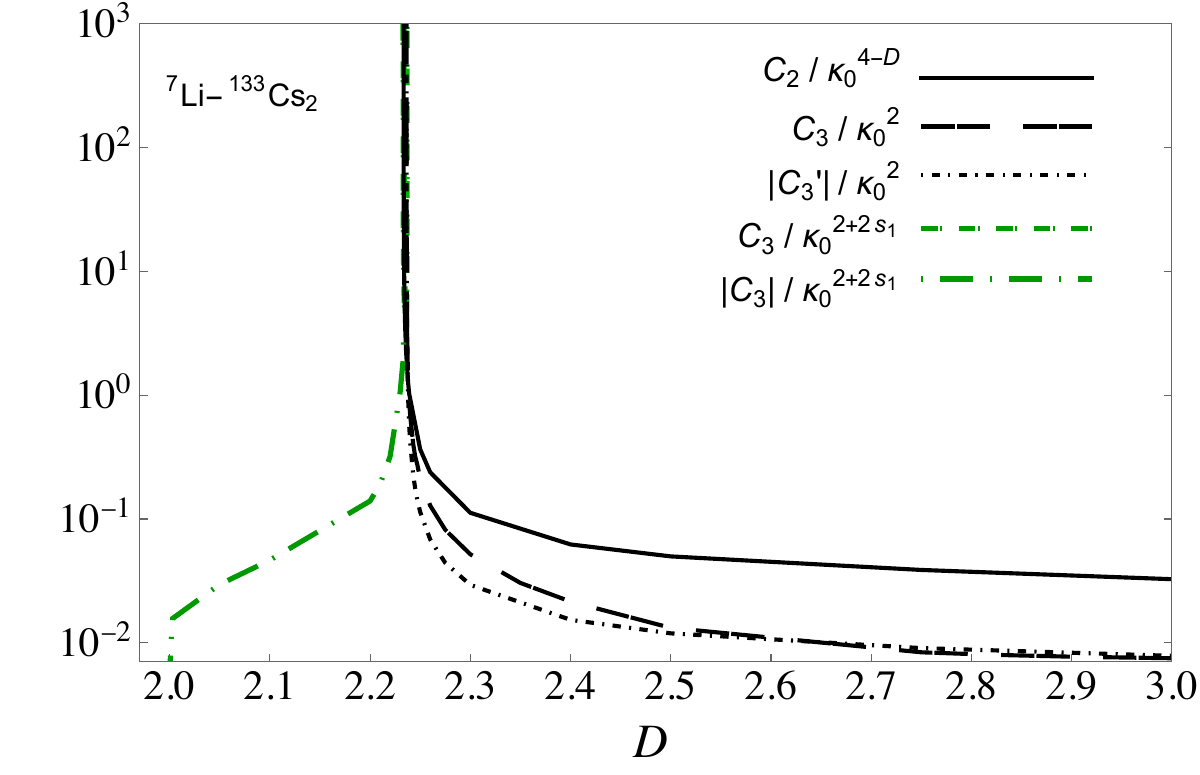}
\includegraphics[width=8.5cm]{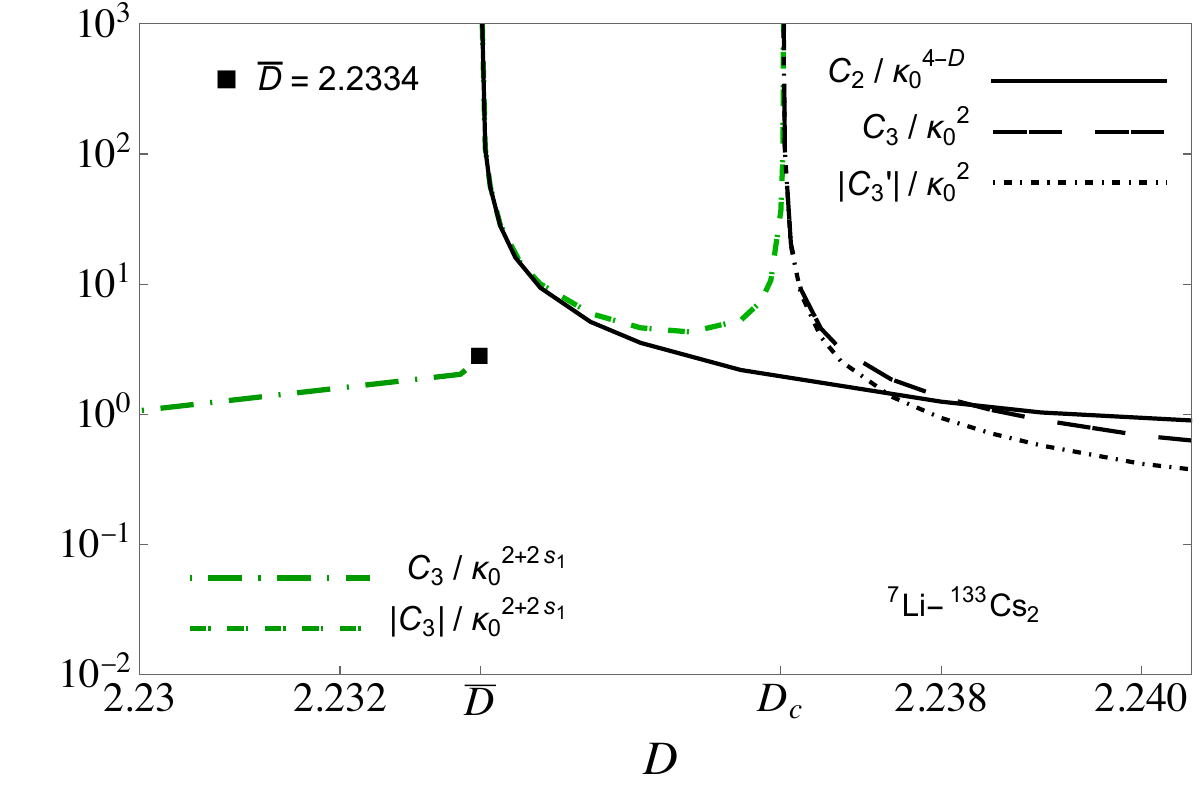}
\caption{The two- and three-body contact parameters $C_2/\kappa_0^{4-D}$, $|C_3'|/\kappa_0^{2}$, and $C_3/\kappa_0^{2}$ as a function of the noninteger dimension~$D$. Left panels: from $D=2$ to $D=3$. Right panels: zoom in around the unatomic critical dimension.} 
\label{fig6}
\end{figure*}

Figure~\ref{fig6} displays the two- and three-body contact parameters as a function of~$D$. They are computed from the equations in the appendices~\ref{appb} and ~\ref{appd}. In the left panels we show the results over a wide interval, $2 \leq D \leq 3$. One can observe  an increase of the contact parameters as $D$ decreases for the $^{7}$Li$-^{23}$Na$_{2}$, $^{7}$Li$-^{87}$Rb$_{2}$ and $^{7}$Li$-^{133}$Cs$_{2}$ compounds in the Efimov regime. These findings were already reported in Ref.~\cite{Nb_AAB_Ddim_Efimov} for a mass-imbalanced Efimov state in the $^6$Li-$^{133}$Cs$_2$ system. In the right panels, we zoom into the narrow region $\overline{D}<D<D_c$. At $\overline{D}$, as discussed in the previous sub-section, there occurs an exchange between the leading and sub-leading contributions in the asymptotic expansion of the single particle momentum distribution. Below $\overline{D}$, where only the continuous scaling term is present, the systems behave as unatomic states. Comparing the results for the $^{7}$Li$-^{23}$Na$_{2}$, $^{7}$Li$-^{87}$Rb$_{2}$, and $^{7}$Li$-^{133}$Cs$_{2}$ compounds, we can observe that the interval between $\overline{D}$ and $D_c$ contracts as the mass imbalance increases.

%%%%%%%%%%%%%%%%FIGURE6%%%%%%%%%%%%%%%%%%%
%%%%%%%%%%%%%%%%%%%%%%%%%%%%%%%%%%%%%%%%%%%%%

\section{CONCLUSIONS}
\label{section4}

Using the Faddeev equations in conjunction with the Bethe-Peierls boundary condition for an infinite scattering length, we studied the behavior of resonant mass-imbalanced three-body system of the type AAB in different regimes according to the noninteger dimension in which the system is embedded. These regimes can be classified as Efimov and unatomic, each presenting different features concerning universal phenomena. The transition between such regimes is made by considering that the system is trapped, with the continuous spatial change of the trap being mimicked by the variation in the noninteger dimension. 

We computed the contact parameters considering the masses configurations of the compounds $^7$Li$-^{23}$Na$_2$, $^7$Li$-^{87}$Rb$_2$ and $^7$Li$-^{133}$Cs$_2$. Starting from $3D$, we found that the three-body parameters tend to increase considerably in magnitude until there is a sharp divergence at the critical dimension. The two-body parameter, although also increasing with decreasing non-integer dimension, was found to be finite at the critical one, diverging at the dimension $\overline{D}$, where the wave function assumes power-law behavior, which is the signature of unatomic states. The narrow spatial region ($\overline{D}<D<D_c$), compared to three identical particles~\cite{unatomic}, becomes smaller as the mass imbalance increases.

%%%%%%%%%%%%%%%%%%%%%%%%%%%%%%%%%%%%%
\section*{ACKNOWLEDGMENTS}
This study was financed, in part, by the São Paulo Research Foundation (FAPESP), Brasil. [grant numbers 2019/07767-1 (T.F.), 
2023/02261-8 (D.S.R.), 2018/25225-9 (G.K.) and 2023/08600-9 (R. M. F.)], Conselho Nacional de Desenvolvimento 
Cient\'{i}fico e Tecnol\'{o}gico (CNPq) [grant numbers 306834/2022-7 (T.F.), 
302105/2022-0 (M.T.Y.) and 309262/2019-4 (G.K.)] and Coordenação de Aperfeiçoamento de Pessoal de Nível Superior (CAPES) Grant number
88887.928099/2023-00 (D.S.R).  This work is a part of the
project Instituto Nacional de  Ci\^{e}ncia e Tecnologia - F\'{\i}sica
Nuclear e Aplica\c{c}\~{o}es  Proc. No. 464898/2014-5.

\appendix

\section{D-dimensional mass imbalanced trimer state}
\label{appa}

In this appendix, we review the derivation of the $D$-dimensional three-body wave function of a mass-imbalanced system at unitarity. We found the solution of the energy 
eigenvalue equation in configuration and in momentum space. The pairs are considered to interact by mean of zero-range potentials, so that the Bethe-Peierls boundary condition is suitable to obtain the trimer scale parameter. 

\subsection{Position space}
\label{appconfspace}

We consider three different particles with masses $m_i$, $m_j$, $m_k$, and coordinates $\textbf{x}_{i}$, $\textbf{x}_{j}$ and $\textbf{x}_{k}$. Ignoring the dynamics of the center of mass, one can describe the system in terms of two relative Jacobi coordinates
\begin{equation}
 \mbox{\boldmath$r$}_{i} = \textbf{x}_{j} - \textbf{x}_{k}\quad\text{and}\quad
 \mbox{\boldmath$\rho$}_{i} = \textbf{x}_i - \frac{m_j\textbf{x}_j+m_{k}\textbf{x}_k}{m_j + m_k} \, ,
\end{equation}
where ($i, j, k$) are taken cyclically among each other. By means of the Faddeev decomposition, one can write the three-body wave function as a sum of three components, which reads
\begin{equation}
\Psi(\textbf{x}_{i},\textbf{x}_{j},\textbf{x}_{k}) = 
\psi^{(i)}(\mbox{\boldmath$r$}_i,\mbox{\boldmath$\rho$}_i) + 
\psi^{(j)}(\mbox{\boldmath$r$}_j,\mbox{\boldmath$\rho$}_j)
+ \psi^{(k)}(\mbox{\boldmath$r$}_k,\mbox{\boldmath$\rho$}_k)\,. 
\end{equation}
Each of these Faddeev components satisfy the free
Schr\"{o}dinger equation.

For convenience, the relative coordinates are rescaled as $
 \mbox{\boldmath$r$}'_{i} = \sqrt{\eta_i}\, \mbox{\boldmath$r$}_{i}\quad$ and $\quad
 \mbox{\boldmath$\rho$}'_{i} = \sqrt{\mu_i}\, \mbox{\boldmath$\rho$}_{i}$, being the three sets of 
 primed coordinates related to each other by the orthogonal transformations
\begin{eqnarray}
\mbox{\boldmath$r$}'_{j}& =& - \mbox{\boldmath$r$}'_{k}\cos\theta_i + \mbox{\boldmath$\rho$}'_{k}
\sin \theta_i, \nonumber \\
\mbox{\boldmath$\rho$}'_{j}& =& - \mbox{\boldmath$r$}'_{k}\sin\theta_i - 
\mbox{\boldmath$\rho$}'_{k}\cos \theta_i,
\end{eqnarray}
where $\tan \theta_i = \left[m_i M/(m_j\ m_k)\right]^{1/2}$, with $M = m_1 + m_2 + m_3$ and the reduced masses are given by
 $\eta_{i} = m_{j}m_{k}/(m_{j}+m_{k})$ and $ \mu_{i} = {m_{i}(m_{j}+m_{k})}/({m_{i}+m_{j}+m_{k}})$.

For three distinct bosons in a state with vanishing total angular momentum, 
one can define a reduced Faddeev component as $
\chi^{(i)}_0 (r'_{i}, \rho'_i) = \left( r'_{i} \ 
\rho'_{i}\right)^{ (D-1)/2} \psi^{(i)}(r'_i,\rho'_i)$.  
To obtain the solutions of the eigenvalue equation for $\chi^{(i)}_0$, one can use the hyperspherical coordinates $r'_i = R \sin \alpha_i$ and $\rho'_i = R \cos \alpha_i $ to separate variables, so that one can write $
\chi^{(i)}_0(R,\alpha_{i})  = \mathcal{C}^{(i)} F(R)\,G^{(i)}(\alpha_{i})$,
where $R^{2}= r_i^{\prime 2}+\rho_i^{\prime2}$, $\alpha_i = \arctan(r'_i/\rho'_i)$ and the coefficients $\mathcal{C}^{(i)}$ represent the weight between the different Faddeev components for mass imbalanced systems. The functions
$F(R)$ and $G^{(i)}(\alpha_{i})$ satisfy the differential equations
\begin{eqnarray}
\left[- \frac{\partial^{2}}{\partial R^{2}} + \frac{s_{n}^{2}-1/4}{R^{2}} + 2\kappa_0^2
\right]\sqrt{R}F(R)=0
 \label{radialwavefunc}
\end{eqnarray}

and

\begin{eqnarray}
\left[- \frac{\partial^{2}}{\partial \alpha_i^{2}} -s_{n}^{2}+\frac{(D-1)(D-3)}{ \sin^2 2 
\alpha_i}\right] G^{(i)}(\alpha_i)=0, 
\label{Eq:angularD}
\end{eqnarray}
where $-\kappa_0^2 = E_3$, being $E_3$ the energy eigenvalue and $s_n$ the Efimov parameter.

Changing variables as $z = \cos 2\alpha_i$ and writing a reduced form of the equation $G^{(i)}(z) = (1-z^2)^{1/4} g^{(i)}(z)$,
Eq.~\eqref{Eq:angularD} becomes the known associated Legendre differential 
equation~\cite{legendrebook} with analytical solutions
\begin{eqnarray} 
G^{(i)}(\alpha_i) &=& \sqrt{\sin2 \alpha_i}\Big[ P_{s_n/2-1/2}^{D/2-1}\,(\cos2\alpha_i) \nonumber \\
&-& \frac{2}{\pi}\tan\big[\pi(s_{n} -1)/2\big] Q_{s_n/2-1/2}^{D/2-1}\,(\cos2\alpha_i)\Big],\ \ \ \ \
\label{Eq:AngSol}
\end{eqnarray}
where $P_{n}^{m}(x)$ and $Q_{n}^{m}(x)$ are the associated Legendre functions. A finite value for the Faddeev component $\psi^{(i)}$ at $\rho_i =0$ imposes that $G^{(i)}(\alpha_i=\pi/2)= 0$, since $\rho_i' = R \cos{\alpha_i}$.

\begin{widetext}

Considering the solution of the hyperradial,
equation~\eqref{radialwavefunc}, and the hyperangular eigenfunction, 
Eq.~\eqref{Eq:AngSol}, each Faddeev component of the wave function is
written as~\cite{betpeiPRA}:
\begin{eqnarray}
\psi^{(i)}(r'_i,\rho'_i) &=&\mathcal{C}^{(i)}   \frac{ K_{ s_n}\left(\sqrt{2} \kappa_0 \sqrt{  r'^{2}_{i}+  \rho'^{2}_{i} } 
\right) }
{ \big(  r'^{2}_{i}+ \rho'^{2}_{i} \big)^{D/2-1/2}}\frac{\sqrt{\sin\big[2 \arctan\left( 
r'_i/\rho'_i\right)\big]}}{\big\{\cos\big[ \arctan\left( r'_i/\rho'_i\right)\big]\ \sin\big[ \arctan\left( 
r'_i/\rho'_i\right)\big]\big\}^{D/2-1/2}}
\nonumber \\
&\times&\left[ P_{s_n/2-1/2}^{D/2-1}\Big\{\cos\big[2 \arctan( 
r'_i/\rho'_i)\big]\Big\}-\frac{2}{\pi}\tan\big[\pi(s_n-1)/2 \big] Q_{s_n/2-1/2}^{D/2-1}\Big\{\cos\big[2 
\arctan( r'_i/\rho'_i)\big]\Big\}\right]\, ,
\label{wavefunction}
\end{eqnarray}
where $K_{ s_n}$ is the modified Bessel function of the second kind. 
\end{widetext}

The BP boundary condition applies
to the total wave function when each relative distance between two of the particles tends to zero. For a chosen pair of coordinates
$(\mbox{\boldmath$r$}_i,\mbox{\boldmath$\rho$}_i)$,
it reads in the unitary limit $a\rightarrow \infty$ as 
\begin{equation}
\label{eq:BP3B}
\hspace{-0.2cm}\left[\frac{\partial}{\partial r_i}  
r_{i}^{\frac{D-1}{2}}\Psi(\mbox{\boldmath$r$}_i,\mbox{\boldmath$\rho$}_i)
\right]_{r_i\rightarrow 0} = \frac{3-D}{2} 
\left[\frac{\Psi(\mbox{\boldmath$r$}_i,\mbox{\boldmath$\rho$}_i)}
{r_{i}^{\frac{3-D}{2}}}\right]_{r_i\rightarrow 0}.
\end{equation}

\begin{widetext}
Taking the three cyclic permutations of $\{i,j,k\}$, one obtain a homogeneous linear system 
\begin{equation} 
\frac{\mathcal{C}^{(i)}}{2}
\left[ \left(\cot\alpha_i\right)^{\frac{D-1}{2}} 
\left( \sin2\alpha_i \frac{\partial}{\partial \alpha_i} 
+ D-3\right) G^{(i)} (\alpha_i) \right]_{\alpha_i\rightarrow 0}
+ (D -2) \left[ \frac{\mathcal{C}^{(j)} \, G^{(j)}(\theta_k)}
{\left(\sin\theta_k \cos\theta_k\right)^{\frac{D-1}{2}}} 
+ \frac{\mathcal{C}^{(k)} \, G^{(k)}(\theta_j)}
{\left(\sin\theta_j \cos\theta_j\right)^{\frac{D-1}{2}}} 
\right] = 0\,,
\label{BPsystem}
\end{equation} 
for $i\neq j \neq k$. The Efimov parameter and the weight between different Faddeev components are obtained by solving the characteristic transcendental equation of the system. 
\end{widetext}

\subsection{Momentum space} 
\label{appmomspace}

We review the derivation of the 
three-body wave function of a mass-
imbalanced system at unitarity in momentum space. For that, we consider $\textbf{k}_i$, $\textbf{k}_j$ and $\textbf{k}_k$ as the momenta of each particle in the rest frame. The Jacobi momenta of the pairs and from one particle to the center of mass of the other two are given, respectively, by
\begin{eqnarray}
\frac{\textbf{p}_i}{\eta_i} = \frac{\textbf{k}_j}{m_j} - \frac{\textbf{k}_k}{m_k}\ \ \text{and}\ \
\frac{\textbf{q}_i}{\mu_i} = \frac{\textbf{k}_i}{m_i}-\frac{\textbf{k}_k+\textbf{k}_k}{m_j+m_k}, \ \ 
\end{eqnarray}
where $(i,j,k)$ are taken cyclically. The Faddeev decomposition in the momentum space reads
\begin{equation}
\label{A11}
\Psi(\textbf{k}_{i},\textbf{k}_{j},\textbf{k}_{k}) = \frac{\chi^{(i)}(\textbf{q}_i)+\chi^{(j)}(\textbf{q}_j)+\chi^{(k)}(\textbf{q}_k)}{E_3 + H_0}, 
\end{equation}
where $H_0$ is the free Hamiltonian in momentum space for any chosen pair of coordinates. The Jacobi momenta $\textbf{p}_i$ and $\textbf{q}_i$ related to each other as
\begin{eqnarray}
    \textbf{q}_j &=& \textbf{p}_i + \frac{m_j}{m_j+m_k}\textbf{q}_i,\nonumber \\
    \textbf{q}_k &=& \textbf{p}_i - \frac{m_k}{m_k+m_j}\textbf{q}_i.
\end{eqnarray}

In order to write the total bound trimer wave function we need to compute the spectator function, $\chi^{(i)}(\textbf{q}_i)$. For that, we need the asymptotic form of 
the Faddeev wave function, Eq.~\eqref{wavefunction}, which is written as
\begin{eqnarray}
\psi^{(i)}\left(\rho'_i,r'_i \right)&\underset{r'_i\to 0}{=}&\mathcal{C}^{(i)} \frac{\sqrt{2} \left[1+\cot \left(D \pi/ 2\right) \cot \left(s_n \pi/2  \right)\right]}{\Gamma \left(2-D/2\right)} \nonumber \\
&\times &\ {r'_i}^{2-D}\frac{K_{s_n}\left(\sqrt{2}\kappa_0 \rho'_{i} \right)}{\rho'_i}, 
\label{wfrlimit}
 \end{eqnarray}
where $\Gamma(z)$ is the gamma function defined for all complex numbers $z$, 
except for the non-positive integers. This condition restricts the validity of our results to the interval $2\leq D <4$. 

The spectator function, namely $B^{(i)}(\rho_i)$, can be found by taking into consideration that each Faddeev component $\psi^{(i)}(\rho'_i,r'_i)$ obeys the Schr\"odinger’s equation with contact interactions, written as
\begin{equation}
\left[\nabla_{r'_i}^{2}+\nabla_{\rho'_i}^{2}- 2\kappa_0^2 \right] \psi^{(i)}(r'_i,\rho'_i)= \delta(r'_i)B^{(i)}(\rho'_i)\,.
\label{freeschroe}
\end{equation}
We substitute Eq.~\eqref{wfrlimit} in~\eqref{freeschroe}, which, in the limit $r'_{i}\to0$, is given by
\begin{eqnarray}
B^{(i)}(\rho'_i)&=&\mathcal{C}^{(i)}\frac{2^{3/2} \pi^{D/2}\left[1+\cot \left(D\pi/ 2\right) \cot \left(s_n \pi/2  \right)\right]}{\Gamma(D/2)\Gamma \left(2-D/2\right)} \nonumber \\
&\times&\frac{K_{ s_n}\left(\sqrt{2}\kappa_0 \rho'_{i} \right)}{\rho'_i}.
 \label{specfuncrho}
\end{eqnarray}
Taking the $D$-dimensional FT
\begin{equation}
\int d^{D}\rho'_i \exp(-i \textbf{q}'_i.\bm{\rho}'_i)B^{(i)}(\rho'_i) = \chi^{(i)}(q'_i)\,,
\end{equation}
one can write the spectator function in momentum space ($q'_i=  q_i/\sqrt{\mu_i}$) as
 \begin{eqnarray}
 \chi^{(i)}(q'_i)&=&\mathcal{C}^{(i)}\mathfrak{F}_{(D,s_n)}
\kappa_{0}^{1-D} \nonumber \\
&\times&
H_2  \tilde{F}_1 \left(\mathcal{F}_{(D,s_n)}^{-} ,\mathcal{F}^{+}_{(D,s_n)},\frac{D}{2},-\frac{q_i^{\prime 2}}{2\kappa_0^2}
 \right),\ \ 
 \label{regularspec}
\end{eqnarray} 
where
\begin{eqnarray}
\mathfrak{F}_{(D,s_n)}& \equiv & \frac{2^{1+D/2} \pi^{D-1}\  \Gamma \big[ \mathcal{F}_{(D, s_n)}^{+}\big]\,
 \Gamma \big[ \mathcal{F}_{(D,s_n)}^{-}\big] }{(2-D)}\nonumber \\
 &\times&\cos \left[\frac{\pi}{2}(D-s_n)  \right]\csc \left( \frac{\pi}{2} s_n\right)
 \label{DefF}
\end{eqnarray}
and $H_2 \tilde{F}_1(a,b,c,z)$ is the regularized hyper-geometrical function with $\mathcal{F}_{(D, s_n)}^{\pm} \equiv (D-1\pm s_n)/2$.

\section{Momentum distribution tail at the Efimov region $D_c <D \leq 3$}
\label{appb}

In this appendix, in order to be self contained, we review the derivation of the oscillatory and non-oscillatory contributions of the asymptotic single particle momentum distribution in the Efimov regime for systems of the form AAB. 

\subsection{\texorpdfstring{Sub-leading contributions to $n_1(q_B)$} {Sub}}
\label{appn1efimov} 

For large momentum, we use the asymptotic spectator function, Eq.~\eqref{asympspecefimov}. Simple manipulations allow us to separate the oscillatory term in equation~\eqref{eq:n1desenvolved}, namely log-periodic one
\begin{eqnarray}\label{eq:A3}
 n^\text{osc}_{1}(q_B)&=&  \frac{\left| \mathcal{C}^{(B)}\right|^{2}}{q_B^{D+2}}\cos\left[2s_0 \ln\left( \frac{q_B}{\sqrt{2\mu_B}\kappa_0^{*}}\right)\right] |\mathfrak{F}_{(D,s_0)}|^2 \nonumber \\ 
 &\times&\mathcal{S}_{D}\pi\left(1-\frac D2\right)\left[\operatorname{Re}(\mathcal{G})^2+\operatorname{Im}(\mathcal{G})^2\right] \nonumber \\
 &\times& (2\mu_B)^{1+D/2}\csc\left( \frac{D\pi}{2} \right) ,
 \end{eqnarray}
 and the non-oscillatory part
  \begin{eqnarray} \label{eq:A4}
 n^\text{nosc}_{1}(q_B)&=&  \frac{\left| \mathcal{C}^{(B)}\right|^{2}}{q_B^{D+2}}|\mathfrak{F}_{(D,s_0)}|^2 \mathcal{S}_{D}\pi \left(\frac{2-D}{2}\right)(2\mu_B)^{1+D/2}\nonumber \\
 &\times&\left[\operatorname{Re}(\mathcal{G})^2+\operatorname{Im}(\mathcal{G})^2\right] \csc\left( \frac{D\pi}{2} \right)\, ,
 \end{eqnarray}
where $\mathcal G$ is written in Eq.~\eqref{eq:G}.
 
\subsection{\texorpdfstring{Sub-leading contributions to $n_2(q_B)$}{Sub}}
\label{appn2efimov}

To derive the oscillatory part of $n_2(q_B)$, we insert the large momenta spectator function, Eq.~\eqref{asympspecefimov}, in the subtracted Eq.~\eqref{eq:n2desenvolved}, i.e.,
$n_{2}(q_B) -C_2/q_B^{4}$. Using trigonometric identities for $\cos^{2}(x)$, one can perform some manipulations 
\begin{eqnarray}
\cos^{2}\left[s_0 \ln\left({\frac{q'_{A}\ q_B}{\sqrt{2\mu_A}\kappa_0^{*}}}\right)\right]&=&\frac{1}{2}\nonumber \\
&+&\frac{1}{2}\bigg\{\cos\left[2s_0 \ln\left({\frac{ q_B}{\sqrt{2\mu_A}\kappa_0^{*}}}\right)\right]\nonumber \\
&\times&\cos\left[2s_0 \ln(q'_{A})\right]\nonumber \\
&-&\sin\left[2s_0 \ln\left(\frac{ q_B}{\sqrt{2\mu_A}\kappa_0^{*}}\right)\right]\nonumber \\
&\times&\sin\left[2s_0 \ln({q'_{A})}\right]\bigg\},
\end{eqnarray}
so that the oscillatory term can be written as
\begin{eqnarray}\label{eq:B7}
n^\text{osc}_{2}(q_B) -\frac{C_2}{q_B^{4}}
&=&\frac{\left| \mathcal{C}^{(A)}\right|^{2}}{q_B^{D+2}}\frac{2^{1+D}\mathcal{S}_D}{\mu_A^{1-D}}\left[\operatorname{Re}(\mathcal{G})^2+\operatorname{Im}(\mathcal{G})^2\right]\nonumber \\
&\times&|\mathfrak{F}_{(D,s_0)}|^2\int^\infty_0 dq'_A \ q_A^{\prime 1-D}   \nonumber \\
&\times&\left( \mathcal{H}(q'_A)-\frac{4 \mathcal{A}^2}{(\mathcal{A}+1)^2}\right)\nonumber \\
&\times&\bigg\{\cos\left[2s_0 \ln\left({\frac{ q_B}{\sqrt{2\mu_A}\kappa_0^{*}}}\right)\right]\nonumber \\
&\times&\cos\left[2s_0 \ln(q'_{A})\right]\nonumber \\
&-&\sin\left[2s_0 \ln\left(\frac{ q_B}{\sqrt{2\mu_A}\kappa_0^{*}}\right)\right]\nonumber \\
&\times&\sin\left[2s_0 \ln({q'_{A})}\right] \bigg\},
\end{eqnarray}
while the non-oscillatory one reads
\begin{eqnarray}\label{eq:B8}
n^\text{nosc}_{2}(q_B)  -\frac{C_2}{q_B^{4}}
&=&\frac{\left| \mathcal{C}^{(A)}\right|^{2}}{q_B^{D+2}}\frac{2^{1+D}\mathcal{S}_D}{\mu_A^{1-D}}\left[\operatorname{Re}(\mathcal{G})^2+\operatorname{Im}(\mathcal{G})^2\right] \nonumber \\
&\times&|\mathfrak{F}_{(D,s_0)}|^2\int^\infty_0 dq'_A \ q_A^{\prime 1-D}  \nonumber \\
&\times&\left( \mathcal{H}(q'_A)-\frac{4 \mathcal{A}^2}{(\mathcal{A}+1)^2}\right)\,.\ \ \ 
\end{eqnarray}
 
\subsection{\texorpdfstring{Sub-leading contributions to $n_3(q_B)$}{Sub}}
\label{appn3efimov}

The asymptotic form is found by using the spectator function from Eq.~\eqref{asympspecefimov} and computing the integral in Eq.~\eqref{eq:n3desenvolved}, leading to
\begin{eqnarray}
 n_{3}(q_B)&=&\frac{\mathcal{C}^{(B)^{\scalebox{0.7}{*}}}\mathcal{C}^{(A)} }{q_B^{D+2}} 2^{D+3}\mathcal{S}_D   (\mu_B\mu_A)^{D/2-1/2}\nonumber \\
 &\times&\left[\operatorname{Re}(\mathcal{G})^2+\operatorname{Im}(\mathcal{G})^2\right]\left\{\cos\left[s_0 \ln\left(\frac{ q_B}{\sqrt{2\mu_B}\kappa_0^{*}}\right)\right]\right.\nonumber \\
 &\times&\left.\cos\left[s_0 \ln\left(\frac{ q_B}{\sqrt{2\mu_A}\kappa_0^{*}}\right)\right]\right.\nonumber \\
 &\times&\left.\int^\infty_0 dq'_A  \mathcal{H}(q'_A)\cos\left[s_0 \ln\left(q'_{A}\right)\right]\right. \nonumber \\
 &-&\left.\sin\left[s_0 \ln\left(\frac{ q_B}{\sqrt{2\mu_B}\kappa_0^{*}}\right)\right]\sin\left[s_0 \ln\left(\frac{ q_B}{\sqrt{2\mu_A}\kappa_0^{*}}\right)\right]\right.\nonumber \\
 &\times&\left.\int^\infty_0 dq'_A  \mathcal{H}(q'_A)\sin\left[s_0 \ln\left(q'_{A}\right)\right]\right\}\,.
 \end{eqnarray} 
The algebraic manipulation of the cosines and sines in the above equation allows us to identify the oscillatory term as
\begin{eqnarray}\label{eq:C4}
 n^\text{osc}_{3}(q_B)&=&\frac{\mathcal{C}^{(B)^{\scalebox{0.7}{*}}}\mathcal{C}^{(A)}  }{q_B^{D+2}} 2^{D+2}\mathcal{S}_D   (\mu_B\mu_A)^{D/2-1/2}  \nonumber \\
 &\times &|\mathfrak{F}_{(D,s_0)}|^2\left[\operatorname{Re}(\mathcal{G})^2+\operatorname{Im}(\mathcal{G})^2\right] \nonumber \\
 &\times&\left\{\cos\left[s_0 \ln\left(\frac{ q_B^{2}/\kappa_0^{*2}}{2\sqrt{\mu_A\mu_B}}\right)\right]\right.\nonumber \\
 &\times&\left.\int^\infty_0 dq'_A  \mathcal{H}(q'_A)\cos\left[s_0 \ln\left(q'_{A}\right)\right]\right. \nonumber \\
 &-&\left.\sin\left[s_0 \ln\left(\frac{ q_B^{2}/\kappa_0^{*2}}{2\sqrt{\mu_A\mu_B}}\right)\right]
 \right.\nonumber \\
 &\times&\left.\int^\infty_0 dq'_A  \mathcal{H}(q'_A)\sin\left[s_0 \ln\left(q'_{A}\right)\right]\right\}, 
 \end{eqnarray}
 and the non-oscillatory one as
 \begin{eqnarray}\label{eq:C5}
 n^\text{nosc}_{3}(q_B)&=&\frac{\mathcal{C}^{(B)^{\scalebox{0.7}{*}}}\mathcal{C}^{(A)}  }{q_B^{D+2}} 2^{D+2}\mathcal{S}_D   (\mu_B\mu_A)^{D/2-1/2} \nonumber \\ 
 &\times& |\mathfrak{F}_{(D,s_0)}|^2\left[\operatorname{Re}(\mathcal{G})^2+\operatorname{Im}(\mathcal{G})^2\right] \nonumber \\
 &\times&\left\{\cos\left[s_0 \ln\left(\sqrt{\frac{\mu_B}{\mu_A}}\right)\right]\right.\nonumber \\
 &\times&\left. \int^\infty_0  dq'_A  \mathcal{H}(q'_A)\cos\left[s_0 \ln\left(q'_{A}\right)\right]\right. \nonumber \\
 &-&\left.\sin\left[s_0 \ln\left(\sqrt{\frac{\mu_B}{\mu_A}}\right)\right] \right. \nonumber \\
 &\times&\left.\int^\infty_0 \hspace{-.2cm} dq'_A  \mathcal{H}(q'_A)\sin\left[s_0 \ln\left({q'_{A}}\right)\right]\right\}. \nonumber \\
 \end{eqnarray}

\subsection{\texorpdfstring{Sub-leading contributions to $n_4(q_B)$}{Sub}}
\label{appn4efimov}

The product of the spectator functions, Eq.~(\ref{asympspecefimov}), allow us to write
\begin{eqnarray}
 \chi^{(A)}\overset{*}{(}q_B \ p'_{B -})&.&\chi^{(A)}(q_B \ p'_{B +}) = |\mathcal{C}^{(A)}|^2 2|\mathfrak{F}_{(D,s_0)}|^2 \nonumber \\
 &\times& \left[\operatorname{Re}(\mathcal{G})^2+\operatorname{Im}(\mathcal{G})^2\right]   \nonumber \\
 &\times&\left(\frac{  q_B}{\sqrt{2\mu_A}}\sqrt{p'_{B-}\ p'_{B+}}\right)^{2-2D}\nonumber \\
 &\times& \left\{ \cos\left[s_0 \ln\left(\frac{  p'_{B+}}{p'_{B-}}\right)\right] \right. \nonumber \\
 &+&\left.\cos\left[s_0 \ln\left(\frac{ q_B^{2}\ p'_{B+}p'_{B-}}{2\mu_A\kappa_0^{*\ 2}}\right)\right]\right\}.
 \end{eqnarray}
Then, the oscillatory and non-oscillatory contribution from Eq.~\eqref{eq:n4desenvolved} is given, respectively, by
\begin{eqnarray}\label{eq:D4}
 n^\text{osc}_{4}(q_B)&=&\frac{|\mathcal{C}^{(A)}|^{2}}{q_B^{D+2}}  \frac{2^{2+D}\pi^{D/2-1/2}}{\Gamma[D/2-1/2]}|\mathfrak{F}_{(D,s_0)}|^2\mu_A^{D-1}\nonumber \\
 &\times&\left[\operatorname{Re}(\mathcal{G})^2+\operatorname{Im}(\mathcal{G})^2\right] \int_0^{\pi}d\theta \sin^{D-2}\theta \nonumber \\ 
 &\times&  \int^\infty_0 dp'_B \frac{p_B^{\prime D-1}  }
 { \left[ p_{B}^{\prime 2} + 1/2\mu_B \right]^{2} }\ \mathcal{W}^{1/2-D/2}  \nonumber \\ 
  &\times&   \cos\left[s_0 \ln\left( \frac{q_B^2}{2\mu_A \kappa_0^{*2}}\, \mathcal{W}^{1/2}\right)\right]\, ,
 \end{eqnarray}
and  
   \begin{eqnarray}\label{eq:D5}
 n^\text{nosc}_{4}(q_B)&=&\frac{|\mathcal{C}^{(A)}|^{2}}{q_B^{D+2}}  \frac{2^{2+D}\pi^{D/2-1/2}}{\Gamma[D/2-1/2]}|\mathfrak{F}_{(D,s_0)}|^2\mu_A^{D-1} \nonumber \\
 &\times&\left[\operatorname{Re}(\mathcal{G})^2+\operatorname{Im}(\mathcal{G})^2\right]\int_0^{\pi}d\theta \sin^{D-2}\theta \nonumber \\ 
  &\times &  \int^\infty_0 dp'_B \frac{p_B^{\prime D-1}  }
 { \left[ p_{B}^{\prime 2} + 1/2\mu_B \right]^{2} }\ \mathcal{W} ^{1/2-D/2} \nonumber \\ 
  &\times &    \cos\left[s_0 \ln\left(  \sqrt{\frac{p^{\prime 2}_B+\frac14+ p'_B\cos\theta}{p^{\prime 2}_B+\frac14- p'_B\cos\theta}} \right)\right]\, ,\hspace{0.7cm}
 \end{eqnarray}
  where 
 \begin{equation}
  \mathcal{W}\equiv\left(p^{\prime 2}_B+\frac{1}{4}+ p'_B\cos\theta\right)\left(p^{\prime 2}_B+\frac{1}{4}- p'_B\cos\theta\right)\, .\nonumber
 \end{equation}

\section{Momentum distribution tail at the unatomic region $2<D<D_c$}
\label{appd}

In this appendix, we present the derivation of the oscillatory and non-oscillatory contributions of the asymptotic single particle momentum distribution in the unatomic regime for mass imbalanced systems of the form AAB.

 \subsection{Sub-leading contributions to $n_1(q_B)$}
\label{appn1una}

For large momentum in the unatomic region, we use the asymptotic spectator function, Eq.~\eqref{asympespectunatomic}, and from simple manipulations we can write Eq.\eqref{eq:n1desenvolved} as
\begin{eqnarray}
 n_{1}(q_B)&=&  \frac{\left| C^{(B)}\right|^{2}}{q_B^{D+2+2s_1}}\kappa_0^{2s_1}\pi(2\mu_B)^{1+s_1+D/2}\mathfrak{F}_{(D,s_1)}^2\mathcal{S}_{D}  \nonumber \\
 &\times&\mathcal{G}_{(-s_1)}^2 \left(\frac{2-D}{2} \right)\csc\left( \frac{D\pi}{2} \right).
 \end{eqnarray}
 
 \subsection{Sub-leading contributions to $n_2(q_B)$}
\label{appn2una}
 
To derive $n_2(q_B)$ at large momentum, we introduce the asymptotic spectator function, Eq.~\eqref{asympespectunatomic}, in the first term of Eq.~\eqref{eq:n2desenvolved}, i.e.,
$n_{2}(q_B) -C_2/q_B^{4}$ to write
\begin{eqnarray}
n_{2}(q_B)  -\frac{C_2}{q_B^{4}}
&=&\frac{\left| C^{(A)}\right|^{2}}{q_B^{D+2+2s_1}}\kappa_0^{2s_1}(2\mu_A)^{D+s_1-1}\mathfrak{F}_{(D,s_1)}^2  \nonumber \\
 &\times&\mathcal{G}_{(-s_1)}^2 2\mathcal{S}_{D}\int^\infty_0 dq'_A \ q_A^{\prime 1-D-2s_1}  \nonumber \\
 &\times&\left( \mathcal{H}(q'_A)-\frac{4 \mathcal{A}^2}{(\mathcal{A}+1)^2}\right)\,.
 \label{n2bar}
\end{eqnarray}
The expression above is divergent for
 $D=\overline D=2-2s_1$ ($\overline D=2.2456$ for  $^7$Li - $^{133}$Rb$_{2}$), and, as long as
\begin{equation}
D>\overline D  \, , 
\end{equation}
which corresponds to $\Delta_{n}=-D-2s_1-2>-4 $, it is finite. The UV divergence at $D=\overline D$ is  due to the term having the $4\mathcal{A}^2/(\mathcal{A}+1)^2 $ in the integral of Eq.~\eqref{eq:n2desenvolved}, while the term with $\mathcal{H}(y)$ is UV finite and due to its expansion around $y=0$
\begin{eqnarray}
    \mathcal{H}(y)&\underset{y  \rightarrow 0}{=}& \frac{4\mathcal{A}^2}{(\mathcal{A}+1)^2 } - 16\mathcal{A}^3\frac{ \mathcal{A}(D+3)+D }{D(\mathcal{A}+1)^4}y^2 \nonumber\\
    &+&\mathcal{O}\left(y^3\right),
\end{eqnarray}
the infrared divergence from the term $4\mathcal{A}^2/(\mathcal{A}+1)^2 $ appearing when approaching $D\to\overline D_+$ is canceled. 

One should note that for $D\to\overline D_+$ ($\overline D=2-2s_1$) the density $n_2(q)$ is finite, although $C_2$  and the contribution to $C_3$ diverges. The reason for the cancellation  can be easily seen by noticing that  $C_2\to -C_3\to \infty$, and thus the divergent contribution of the two-body contact is exactly canceled by the three-body contact, letting the tail of $n_2$ finite.

For $2<D<\overline D$, the formula for $C_2$, Eq.~\eqref{eq:c2}, is not valid because the factorization used to define the two-body contact cannot be done due to the UV divergence, and consistently only remains in $n_2(q)$ the contribution to the three-body contact, which now is the leading one
\begin{eqnarray}
\label{n2unatomic}
n_{2}(q_B)
&=&\frac{\left| C^{(A)}\right|^{2}}{q_B^{D+2+2s_1}}\kappa_0^{2s_n}(2\mu_A)^{D+s_1-1}\mathfrak{F}_{(D,s_1)}^2  \nonumber \\
 &\times&\mathcal{G}_{(-s_n)}^2 2\mathcal{S}_{D}\int^\infty_0 dq'_A \ q_A^{\prime 1-D-2s_1}   \mathcal{H}(q'_A)\,.
\end{eqnarray}

When approaching $D\to \overline D$, the above expression has an IR divergence, so that the calculation of the asymptotic form of $n_2(q)$ has to be performed with
 \begin{eqnarray}
n_{2}(q_B) &= & \frac{2\mathcal{S}_D }{q_B^{4-D}}\int^\infty_0 dq'_A \, q_A^{\prime D-1} | \chi^{(A)}(q_B\, q'_A) |^{2} \nonumber \\
&\times&\mathcal{H}(q'_A),
\label{n2unatomic}
\end{eqnarray}
using equation~\ref{regularspec} which is free of UV and IR divergences. In practice, the asymptotic value of $q^{D+2+2S_1}n_2(q)$ for $D$  approaching $\overline D_-$ appears at larger and larger values of the momentum, and exactly in the limit of $D\to\overline D_-$, one has that $q^{4}n_2(q)$ is finite for $q\to \infty$.

 \subsection{Sub-leading contributions to $n_3(q_B)$}
\label{appn3una}

The asymptotic form is found by using the spectator function from Eq.~\eqref{asympespectunatomic} to compute the expression~\eqref{eq:n3desenvolved}, leading to
\begin{eqnarray}
 n_{3}(q_B)&=&\frac{ C^{(B)^{\scalebox{0.7}{*}}}C^{(A)}}{q_B^{D+2+2s_1}}\kappa_0^{2s_1}\left(\sqrt{2\mu_B}\right)^{D+s_1-1}\mathfrak{F}_{(D,s_1)}^2  \nonumber \\
 &\times&4\mathcal{S}_{D}\left(\sqrt{2\mu_A}\right)^{D+s_1-1}\mathcal{G}_{(-s_1)}^2\nonumber \\
 &\times& \int^\infty_0 dq'_A \ q_A^{\prime\ -s_1}  \ \mathcal{H}(q'_A)\,. 
 \end{eqnarray}

\subsection{Sub-leading contributions to $n_4(q_B)$}
\label{appn4una}

Finally, the asymptotic contribution $n_4$ is found by using the spectator function from Eq.~\eqref{asympespectunatomic} to compute equation~\eqref{eq:n4desenvolved}, leading to
\begin{eqnarray}
 n_{4}(q_B)&=&\frac{\left| C^{(A)}\right|^{2}}{q_B^{D+2+2s_1}}\kappa_0^{2s_1}(2\mu_A)^{D+s_1-1}\mathfrak{F}_{(D,s_1)}^2  \nonumber \\
 &\times&\mathcal{G}_{(-s_1)}^2 \frac{4\pi^{(D-1)/2}}{\Gamma[(D-1)/2]}\int^\infty_0 dp'_B \ \frac{p_B^{\prime D-1}  }
 { \left[ p_{B}^{\prime 2} + 1/2\mu_B\right]^{2} } \nonumber \\
 &\times& \int_0^{\pi}d\theta \sin^{D-2}\theta\left( \sqrt{p^{\prime 2}_B+\frac14+ p'_B\cos\theta} \right)^{1-D-s_1} \nonumber \\
 &\times&\left(\sqrt{p^{\prime 2}_B+\frac14-p'_B\cos\theta}\right)^{1-D-s_1}\,.
 \end{eqnarray}

%%%%%%%%%%%%%%%%%%%%%%%%%%%%%%%%%%%%%%%%%%%%%%%%%%%%%%%%%%%%%%%%%%%%%%

\end{document}